\documentclass[]{elsarticle}
\usepackage[commentmarkup=todo,authormarkup=none,commandnameprefix=always,final]{changes} % add "final" to remove markup

\usepackage[margin=1in]{geometry}
\usepackage[hidelinks]{hyperref}
\usepackage{graphicx}
\usepackage{amsmath,amsthm, amssymb, float, bm, subcaption}
\biboptions{sort&compress}
\ifdefined\usetodonotes
\usepackage{todonotes}

\fi
\usepackage{soul}
\usepackage{tcolorbox}
\definecolor{C0}{HTML}{1F77B4}
\definecolor{C1}{HTML}{FF7F0E}
\definecolor{C2}{HTML}{2ca02c}
\definecolor{C3}{HTML}{d62728}
\definecolor{C4}{HTML}{9467bd}
\definecolor{C5}{HTML}{8c564b}

\definechangesauthor[color=C0]{R1} % changes relevant to R1
\definechangesauthor[color=C1]{R5} % changes relevant to R2
\definechangesauthor[color=C2]{RA} % changes relevant to R2
\makeatletter
\@namedef{Changes@AuthorColor}{C2}
\colorlet{Changes@Color}{C2}
\renewcommand{\Changes@Markup@comment}[3]{%
  \IfStrEq{\Changes@optioncommentmarkup}{todo}%
		{\colorlet{Changes@todocolor}{authorcolor}\todo[color=Changes@todocolor!10, bordercolor=Changes@todocolor, linecolor=Changes@todocolor!70, nolist]{\textbf #1}}{}}
\makeatother
\makeatletter
\def\ps@pprintTitle{%
  \let\@oddhead\@empty
  \let\@evenhead\@empty
  \def\@oddfoot{\reset@font\hfil\thepage\hfil}
  \let\@evenfoot\@oddfoot
}
\makeatother

\usepackage[noabbrev,capitalise]{cleveref}
\newtheorem{theorem}{Theorem}

\title{Self-similar diffuse boundary method for phase boundary driven flow}

\author{\texorpdfstring{Emma M.~Schmidt}{Emma M. Schmidt}}
\author{\texorpdfstring{J.~Matt Quinlan}{J. Matt Quinlan}}
\author{\texorpdfstring{Brandon Runnels\corref{cor1}}{Brandon Runnels}}
\cortext[cor1]{Corresponding author}
\address{Department of Mechanical and Aerospace Engineering, University of Colorado, Colorado Springs, CO USA}

\begin{document}
\begin{abstract}
  %% Note: JFM has a 250 word limit for the abstract.
  Interactions between an evolving solid and inviscid flow can result in substantial computational complexity, particularly in circumstances involving varied boundary conditions between the solid and fluid phases.
  Examples of such interactions include melting, sublimation, and deflagration, all of which exhibit bidirectional coupling, mass\slash heat transfer, and topological change of the solid-fluid interface.
  The diffuse interface method is a powerful technique that has been used to describe a wide range of solid-phase interface-driven phenomena.
  The implicit treatment of the interface eliminates the need for cumbersome interface tracking, and advances in adaptive mesh refinement have provided a way to sufficiently resolve diffuse interfaces without excessive computational cost.
  However, the general scale-invariant coupling of these techniques to flow solvers has been relatively unexplored.
  In this work, a robust method is presented for treating diffuse solid-fluid interfaces with arbitrary boundary conditions.
  Source terms defined over the diffuse region mimic boundary conditions at the solid-fluid interface, and it is demonstrated that the diffuse length scale has no adverse effects.
  To show the efficacy of the method, a one-dimensional implementation is introduced and tested for three types of boundaries: mass flux through the boundary, a moving boundary, and passive interaction of the boundary with an incident acoustic wave.
  These demonstrate expected behavior in all cases.
  Convergence analysis is also performed and compared against the sharp-interface solution, and linear convergence is observed.
   This method lays the groundwork for the extension to viscous flow, and the solution of problems involving time-varying mass-flux boundaries.
\end{abstract}

\ifdefined\usetodonotes
\begin{tcolorbox}[title={\bf Todonote convention}]
  Please use the following convention when making notes:
  \begin{center}\verb-\yourname{Addresseename, we need to XYZ}-\end{center}
  In other words use your name macro to make any comments, then address specific people in the text.
  For instance, if Brandon wants Bob to run more results, he should express this in the following way.
  \begin{center}
    \verb-\brandon{Bob, we need to run more results}-
  \end{center}
\end{tcolorbox}
\listoftodos
\setcounter{page}{0}
\fi

\maketitle

%\begin{keywords}
%\end{keywords}

\section{Introduction}

Many physical problems of interest arise from flow driven by complex solid-fluid boundary interactions. 
Examples include the melting of complex structures (such as snowflakes), the sublimation of dry ice, the development of solid microstructure during metal casting, and the growth of dendrites in batteries.
One particular application of interest is the deflagration of solid composite propellants (SCPs) such as that formed by ammonium perchlorate (AP) particles in a hydroxyl-terminated polybutadiene (HTPB) binder.
The combustion of AP\slash HTPB mixtures is a complex multi-phase problem, where the burn rate is affected by both the properties of the solid (such as oxidizer particle size and distribution) and the gas dynamics of the products~\cite{thomas2019comprehensive}.
In addition, the solid and gas phase evolution are tightly coupled with mass transportation and heat transfer.
Specifically, heat transfer from the combustion to the solid phase causes AP to experience dissociative sublimation (and, away from the binder, deflagration), while the HTPB binder experiences pyrolysis~\cite{cai2008model}.
The result is a highly variable multi-species mass flux across a time-varying boundary with a complex lifted diffusion flame~\cite{murphy1974ammonium}.
Since the flame is nearest to the solid phase near AP\slash HTPB interfaces, resulting in the highest heat transfer rates, the solid phase surface may become highly irregular~\cite{dennis2019combustion}.
In fact, the surface can have large cavities, peninsulas, and even islands of binder that are ejected from the solid phase.
The irregularity, topological changes, and time dependence of the solid-gas boundary results in a number of modeling challenges.

Diffuse interface methods\added{, the phase field method in particular,} have been widely used to model complex interface evolution \added{in the solid phase} \cite{du2020phase,tonks2019phase}.
These methods rely on a surrogate field (e.g., an order parameter) to act as an indicator variable for the domain, which varies smoothly in space and time.
Because diffuse interface methods treat the interface implicitly, rather than explicitly, they are able to handle complex interface geometries and topological changes automatically.
\deleted{A popular diffuse interface method for problems of this type is the phase field method.}
Phase field models generally evolve the order parameter through a kinetic gradient descent differential equation that minimizes a free energy.
Non-local terms in the free energy cause the interface to diffuse over a controllable width, which may or may not have physical significance~\cite{qin2010phase}.  
The phase field method is used in a wide range of phase boundary applications, including growth of dendrites~\cite{karma1998quantitative,wang2020application}, fracture in solid media~\cite{kuhn2010continuum,agrawal2021block}, polycrystalline solidification~\cite{granasy2004modelling,granasy2014phase}, and large deformation of soft materials~\cite{hong2013phase}.
Even nanoscale phase boundaries between atomistic and fluid media have been captured using the phase field crystal method~\cite{mellenthin2008phase,provatas2010phase,berry2008melting}.
Recent work by the authors include the application of the phase field method to the problem of solid composite propellant deflagration \cite{kanagarajan2022diffuse,kanagarajan2022phase}.
Despite its success at modeling the phase boundary evolution, relatively little attention has been paid to the fluid phase.
Some applications have featured concurrent solid and fluid simulations (e.g., dendrite growth under forced flow~\cite{lan2002adaptive,lan2004efficient}), but the focus has been primarily on the passive interaction of the flow with the solid phase, rather than interface-driven flow.
\replaced{Usually, the fluid phase is not resolved, and any interaction across the boundary are treated in a reduced-order manner.}{Usually, however, the fluid phase is not resolved at all in some cases, and any interaction across the boundary are treated in a reduced-order manner.}
An improved coupling mechanism is needed to capture flow driven by a mass flux, specifically for solid composite deflagration.

\added[id=R5,comment={5.1}]{
Interior boundaries are inherent to many fluid dynamic systems, and a wide range of methods have been proposed to treat them.
In cases where the interface evolution is slow, a viable approach is to conform the mesh to the boundary itself, for instance, using the method of boundary fitted coordinates (BFC) \cite{thompson1982boundary,bijl1998unified}.}
Several methods have been previously proposed to model boundary-driven flows with extreme time-varying morphology.
\added[id=R5,comment={5.1}]{The marker-and-cell (MAC) method was proposed as one way to account for complex evolving boundaries of incompressible liquids in an Eulerian frame \cite{welch1965mac,tome1994gensmac,mckee2004recent,mckee2008mac}. Alternatively, }
Lagrangian approaches evolve the mesh with the material, and provide high accuracy boundary evolution~\cite{quan2007moving}; these methods are prone to ``mesh tangling'' or negative cell volumes near complex curvatures.
Similarly, front tracking methods require explicit tracking of the interface \cite{chertock2008interface, sato2013sharp} and may become intractable when topological changes of the boundary occur~\cite{maltsev2022high}.
Alternatively, the level set \added[id=R5,comment={5.1}]{(LS)} method is designed to cope with topological changes, as it employs the isocontours of a surrogate field to evolve the interface \cite{xu2006level, cottet2008eulerian, legay2006eulerian}.
This too may become numerically taxing, however, as the Hamilton-Jacobi equation which describes the isocontours requires high-fidelity numerical methods.
These methods also effectively eliminate the primary advantage of diffuse boundary methods, replacing implicit interface tracking with explicit tracking, and re-introducing the associated difficulties associated with discrete interface resolution.
Moreover, they preclude the treatment of interface diffusivity as a physical parameter (e.g., thickened flame models). 
Consequently, there is a need for a diffuse interface boundary coupling mechanism that is general, unaffected by the diffuse length scale, implicit, and convergent to the sharp interface limit as the diffuse length approaches zero.

Diffuse interface methods have been used extensively in fluid applications~\cite{anderson1998diffuse}, albeit for different applications than those considered here.
One common use of diffuse interfaces is in simulating interaction between two immiscible fluids.
\added[id=R5,comment={5.1}]{
Examples of this include the immersed boundary method (IBM), which is a diffuse method yet still requires explicit interface tracking \cite{peskin1972flow,sotiropoulos2014immersed,griffith2020immersed}.
Another example is the volume-of-fluid method (VOF), which can handle topological transitions but not arbitrary phases or boundary condition types \cite{maric2020unstructured,niethammer2019extended,fuster2018momentum}.
}
Phase field methods have been widely used to capture various types of multi-species or multi-phase flows \cite{fakhari2010phase,soligo2019mass,wang2019brief}, where the order parameter represents a mass or volume fraction.
The phase field framework provides a closure model, as in the  five-equation model of Allaire, Clerc, and Kokh~\cite{allaire2002five}.
In other similar models~\cite{abgrall1996prevent, sainsaulieu1995finite, johnsen2012preventing, jain2020conservative}, the order parameter is subject to an advection-diffusion equation coupled to the  governing equations of the system. 
For a more complicated system such as deflagration to detonation transition, further governing equations are required to capture mass and energy transfer between the phases, as seen in the seven equation model of Baer and Nunziato~\cite{baer1986two}. 
Some diffuse interface approaches in combustion include artificially thickened flame models for premixed combustion, which determine a diffuse length scale based on the criteria that the deflagration speed be consistent with the free boundary problem~\cite{legier2000dynamically, poinsot2005theoretical}.
Using a single order parameter to describe sublimation and pyrolysis does not easily lead to a single physical interpretation, and the number of governing equations required to evolve such an order parameter from first principles would be prohibitively large.
The current work takes a numerical approach, which, while consistent with the models described above, allows for an arbitrarily evolving order parameter.

This phase boundary evolution approach has been applied to solid-fluid interactions in the absence of mass transfer~\cite{wallis2021diffuse,maltsev2022high, kemm2020simple}.
\added[id=R1,comment={1.2}]{Particularly, the works of Jain\cite{jain2021assessment,jain2020diffuseinterface}, Ghaisas\cite{ghaisas2018unified}, Subramaniam\cite{subramaniam2018high}, and Adler\cite{adler2020diffuseinterface} are examples of diffuse interface models for elastic and plastic deformation of solids in compressible flow with immiscible phases.}
This bears some similarity to the wall potential construction used in some cases to restrict fluid motion to the outside of the solid~\cite{favrie2009solid,jacqmin2000contact}.
Localized artificial diffusivity (LAD) methods, which share many characteristics with diffuse interface methods, have been applied to combustion and compressible flows~\cite{lee2017localized,jainstable}. %MQ: tried to make the Jain and Moin cite a little better, but this is not a peer-reviewed source.
Crucially, \deleted[id=R1,comment={1.1}]{preservation of wave speeds and} self-similar behavior is not guaranteed for these formulations. 
For example, self-similarity depends on the form of additional governing equations for the order parameter, and is necessary to ensure that the flow field is not compromised by the addition of the diffuse length scale.
Additionally, it is necessary to develop some mechanism that accounts for more sophisticated solid-fluid boundary conditions at the interface.

The goal of the current work is to devise a formulation of the Euler compressible flow equations that require no discrete boundary conditions, instead accounting for phase boundaries implicitly through diffuse (yet localized) source terms.
The emphasis of this work is on recovering the governing equations and boundary conditions of the classical free boundary problem in the sharp interface limit. 
The evolution of the diffuse interface is effectively treated as an ``input,'' so that only solid to fluid coupling is considered.
The reverse coupling, from the fluid to the solid phase, is already well developed \cite{yu2012extended}, and forms a basis for the theory proposed here.
The generality of the proposed formulation is emphasized, so that minimal constraints are placed on the diffuse boundary, and the prescribed mass\slash momentum\slash energy fluxes are arbitrary (provided they are consistent) \cite{poinsot1992boundary}.

The remainder of this paper is structured as follows: 
Section \ref{sec:formulation} introduces a diffuse interface formulation of the Euler equations with mass, momentum, and energy flux boundary conditions, in the context of a time-evolving diffuse boundary.
We show that the equations limit precisely to the equivalent discrete boundary problem in the sharp interface limit.
We then specialize the formulation to the flow of a thermally and calorically perfect gas.
In \cref{sec:scaling}, we demonstrate that the artificially introduced interface width maintains the characteristic scaling behavior for the flow outside of the interface region.
Conditions for self-similarity inside the diffuse interface are also discussed.
The numerical implementation of the diffuse boundary formulation is briefly described in \cref{sec:implementation}.
\Cref{sec:examples} outlines a suite of numerical examples, in which we demonstrate numerical convergence to the sharp interface limit.
This work concludes in \cref{sec:conclusion} with a discussion of the model results, limitations, and the extension of this approach to more complex numerical work.

\section{Theory}\label{sec:formulation}

\begin{figure}
  \includegraphics[width=\linewidth]{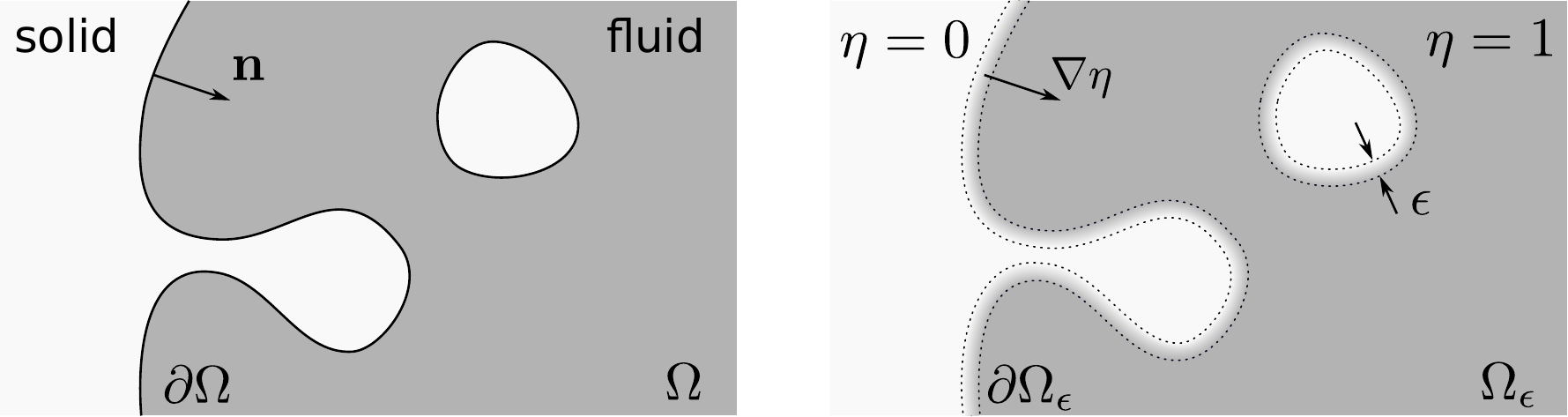}
  \caption{
    Diffuse interface approximation, a discrete boundary (left) is replaced with a smoothly varying field $\eta_\epsilon$ (right). The domain where $\eta_\epsilon=1$ corresponds to the active domain, $\eta_\epsilon=0$ to the inactive domain, and $0<\eta_\epsilon<1$ to the diffuse boundary region, which has a thickness of $\epsilon$.
    The discrete interface is recovered in the limit as $\epsilon\to0$.
    Diffuse thickness may either have physical significance or be considered a numerical regularization only.
  }
  \label{fig:schematic}
\end{figure}

We begin by introducing the general setting for which a diffuse interface model would be applied.
Let $\Omega\subset\mathbb{R}^3$ indicate the domain in which the Euler equations hold (the ``fluid region'') and let $\partial\Omega$ indicate the time-varying boundary of $\Omega$.
The boundary $\partial\Omega$ may exhibit substantial variation over time, generally allowing for complex topological transitions such as interfacial splitting or merging (\cref{fig:schematic} left).
The boundary may act either as a solid wall or as an inlet with arbitrary prescribed flow properties.
No restrictions are made on the geometry of $\Omega$ or its boundary, other than that it be measurable and that the boundary have nonzero radii of curvature everywhere.

The following describes the conservation laws and corresponding boundary conditions associated with Euler flow:
\begin{subequations}\begin{align}
                      \text{Mass conservation:}&&\frac{\partial \rho}{\partial t} + \nabla\cdot (\rho \bm{u})  &= 0  & \bm{x} &\in \Omega \\
                      \text{Momentum conservation:}&&\frac{\partial}{\partial t}(\rho\bm{u})  + \nabla(\rho\,|\bm{u}|^2 + p) &= \bm{0} & \bm{x} &\in \Omega \\
                      \text{Energy conservation:}&&\frac{\partial E}{\partial t} + \nabla\cdot(E+p)\bm{u} &= 0 & \bm{x} &\in \Omega \\
%                      \text{EOS:}&& \frac{1}{2}\rho\,|\bm{u}|^2 + \frac{p}{\gamma-1} &= E & \bm{x}&\in\Omega \\
                      \text{Mass flux BC:}&& \rho\bm{u}\cdot\bm{n} &= \dot m_0 &  \bm{x} &\in \partial\Omega \label{eq:bndry_mass}\\
                      \text{Momentum flux BC:}&& (\rho\bm{u}\cdot\bm{u} + p)\bm{n} &= \dot P_0\bm{n} &  \bm{x} &\in \partial\Omega \label{eq:bndry_momentum}\\
                      \text{Energy flux BC:}&& (E+p)\bm{u}\cdot\bm{n} &= \dot{Q} &  \bm{x} &\in \partial\Omega \label{eq:bndry_energy}.
\end{align}\end{subequations}
An equation of state (EOS) is needed to close the model, but is not of particular importance here as EOS models are, for most fluid applications, entirely local.
The boundary conditions have been presented in terms of abstract mass flux $\dot{m}_0$, momentum flux $\dot{P}_0$, and energy flux $\dot{Q}$.
It is worth noting that boundary conditions described in the form of conserved variable fluxes are not particularly helpful, as it is often impossible to prescribe fluxes in this way.
Moreover, the values on the left hand side of the boundary conditions often contain a mix of known and unknown variables, depending on the boundary type.
In the following sections, the above is specialized to recover the familiar sets of boundary conditions corresponding to the typically imposed boundary types.

\subsection{Diffuse interface construction}

Let $\eta_\epsilon$ be a time-varying, Lipschitz continuous function over $\Omega$ with range $[0,1]$, such that $\eta_\epsilon(\bm{x})=1$ implies that $\bm{x}$ is located in the fluid phase, $\eta_\epsilon(\bm{x})=0$ imply that $\bm{x}$ is located in the solid phase, and that the diffuse boundary region is defined implicitly as $0<\eta_\epsilon(\bm{x})<1$.
The parameter $\epsilon$ determines the diffuse thickness of the boundary, and the sharp interface limit is recovered as $\epsilon\to0$.
Equations governing the evolution of $\eta$, as well as the mass, momentum, and energy fluxes along the diffuse boundary, are available for a wide range of physical systems.
Here, $\eta$ and the corresponding fluxes are taken to be inputs that are supplied either from a concurrent physical diffuse interface model, or by construction.

To simplify the analysis presented below, we consider idealized order parameters only.
Idealized order parameters are defined as those satisfying the property that, for sufficiently small $\epsilon$, the diffuse interface region is restricted to $\partial_\epsilon\Omega=\partial\Omega\times[-\epsilon/2,\epsilon/2]$.
Moreover, in this region,
\begin{align}
  \eta_\epsilon(\bm{y}+s\bm{n}) = \hat\eta_\epsilon(s) \ \ \ \forall\bm{y}\in\partial\Omega, \ \ \ \ s\in[-\epsilon/2,\epsilon/2],
\end{align}
where $\bm{n}$ is the inward-facing normal at $\bm{y}$, $\hat\eta_\epsilon(-\epsilon/2)=0$, $\hat\eta_\epsilon(\epsilon/2)=1$, and $d\hat\eta_\epsilon/d\epsilon$ is a mollifier over $[-\epsilon/2,\epsilon/2]$.
This in turn implies that $\nabla\eta_\epsilon(\bm{y}+s\bm{n}) = \bm{n}\,d\hat\eta_\epsilon/ds$ and $\nabla\eta_\epsilon=\bm{0}$ outside of $\partial_\epsilon\Omega$.

\subsection{Governing equations and boundary condition recovery}
\label{sec:limit}

With the above definitions having been established, it is now possible to construct a formulation for the Euler equations in which the boundary terms (\cref{eq:bndry_mass,eq:bndry_momentum,eq:bndry_energy}) are replaced by source terms, and the domain $\Omega$ is expanded to $\mathbb{R}^3$:
\begin{subequations}\begin{gather}
  \eta_\epsilon\frac{\partial}{\partial t}(\rho) + \nabla\cdot[\eta_\epsilon\rho\bm{u}] = \dot{m}_0|\nabla\eta_\epsilon|  \ \ \ \ \forall\bm{x}\in\mathbb{R}^3, \label{eq:diffuse_mass}\\
  \eta_\epsilon\frac{\partial}{\partial t}(\rho\bm{u}) + \nabla [\eta_\epsilon(\rho\,\bm{u}\cdot\bm{u} + p)] = \dot P_0 \nabla\eta_\epsilon   \ \ \ \ \forall\bm{x}\in\mathbb{R}^3, \label{eq:diffuse_momentum}\\
  \eta_\epsilon\frac{\partial}{\partial t}(E) + \nabla\cdot[\eta_\epsilon(E+p)\bm{u}] = \dot Q_0|\nabla\eta_\epsilon|  \ \ \ \ \forall\bm{x}\in\mathbb{R}^3. \label{eq:diffuse_energy}
\end{gather}\end{subequations}
The remainder of this section will show that the diffuse interface equations recover precisely the original governing equations in the limit as $\epsilon\to0$.
The following theorem will prove useful, the proof of which is supplied in \ref{sec:thm_proof}:
\begin{theorem}\label{thm:boundary_integral}
  Let $\eta_\epsilon$ be an idealized order parameter, and let $f,g$ be vector-valued or scalar-valued bounded functions.
  Then the following holds:
  \begin{align}
    \lim_{\epsilon\to0}\int_A\int_{-\epsilon/2}^{\epsilon/2}\Big[f\eta + g|\nabla\eta|\Big]ds\,dA = \int_A\,g\,dA \ \ \ \ \  \forall A \subset \partial \Omega.
  \end{align}
\end{theorem}

Begin by applying the fundamental lemma of the calculus of variations to \cref{eq:diffuse_mass,eq:diffuse_momentum,eq:diffuse_energy}, converting them to integral (weak) form:
\begin{subequations}\begin{gather}
  \int_B\Big[\eta_\epsilon\frac{\partial}{\partial t}(\rho) + \nabla\cdot[\eta_\epsilon\rho\bm{u}] - \dot{m}_0|\nabla\eta_\epsilon|\Big]\,dV = 0  \ \ \ \ \ \forall B\subset\mathbb{R}^3 \\
  \int_B\Big[\eta_\epsilon\frac{\partial}{\partial t}(\rho\bm{u}) + \nabla[\eta_\epsilon(\rho\,\bm{u}\cdot\bm{u} + p)] - \dot V_0 \nabla\eta_\epsilon\Big]\,dV = 0  \ \ \ \ \ \forall B\subset\mathbb{R}^3, \\
  \int_B\Big[\eta_\epsilon\frac{\partial}{\partial t}(E) + \nabla\cdot[\eta_\epsilon(E+p)\bm{u}]  - \dot Q_0|\nabla\eta_\epsilon|\Big]\,dV = 0 \ \ \ \ \ \forall B\subset\mathbb{R}^3
\end{gather}\end{subequations}
Since the above holds for all $B\subset\mathbb{R}^3$, it must also hold for all $B\subset\Omega_\epsilon$.
Then, because $\eta_\epsilon=1$ and $\nabla\eta_\epsilon=0$ for all $\bm{x}\in B\subset\Omega_\epsilon$, the above reduces to 
\begin{subequations}\begin{gather}
  \int_B\Big[\frac{\partial}{\partial t}(\rho) + \nabla\cdot(\rho\bm{u})\Big]\,dV=0  \ \ \ \ \ \forall B\subset\Omega_\epsilon \\
  \int_B\Big[\frac{\partial}{\partial t}(\rho\bm{u}) + \nabla(\rho\,\bm{u}\cdot\bm{u}) + \nabla p\Big]\,dV=0  \ \ \ \ \ \forall B\subset\Omega_\epsilon \\
  \int_B\Big[\frac{\partial}{\partial t}(E) + \nabla\cdot(E\,\bm{u}) + \nabla\cdot(p\,\bm{u})\Big]\,dV = 0 \ \ \ \ \ \forall B\subset\Omega_\epsilon,
\end{gather}\end{subequations}
In the limit, since $\Omega_\epsilon\to\Omega$, another application of the fundamental lemma recovers the original mass, momentum, and energy balance equations in $\Omega$.
Next, consider the recovery of the boundary equations.
Since the weak form holds for all $B\subset\mathbb{R}^3$, it must also hold for $A\times[-\epsilon/2,\epsilon/2]$, for all $A\subset\partial\Omega$.
Rewriting the integral, applying the product rule, recalling that $\nabla\eta = \bm{n}|\nabla\eta|$, and rearranging terms, allows the integrals to be written as
\begin{subequations}\begin{gather}
  \int_A\int_{-\epsilon/2}^{\epsilon/2}\Big[
  \Big(\frac{\partial}{\partial t}(\rho) + \nabla\cdot(\rho\bm{u})\Big)\eta_\epsilon + \Big(\rho\bm{u}\cdot\bm{n}  - \dot{m}_0\Big)|\nabla\eta_\epsilon|
  \Big]\,ds\,dA
  \\
  \int_A\int_{-\epsilon/2}^{\epsilon/2}\Big[
  \Big(\frac{\partial}{\partial t}(\rho\bm{u}) + \nabla(\rho\,\bm{u}\cdot\bm{u}) + \nabla p\Big)\eta_\epsilon + \Big((\rho\,\bm{u}\cdot\bm{u})\bm{n} + p - \dot P_0 \bm{n}\Big)|\nabla\eta_\epsilon|
  \Big]\,ds\,dA
  \\
  \int_A\int_{-\epsilon/2}^{\epsilon/2}\Big[
  \Big(\frac{\partial}{\partial t}(E) + \nabla\cdot(\,E\,\bm{u}) + \nabla\cdot(p\,\bm{u})\Big)\eta_\epsilon  + \Big(E\,\bm{u}\cdot\bm{n} + p  - \dot Q_0\Big)|\nabla\eta_\epsilon|
  \Big]\,ds\,dA.
\end{gather}\end{subequations}
Taking the limit as $\epsilon\to0$, and applying \cref{thm:boundary_integral}, yields the following boundary integral representation,
\begin{subequations}\begin{gather}
  \int_A\Big[
  \rho\bm{u}\cdot\bm{n}  - \dot{m}_0
  \Big]\,dA=0 \ \ \ \ \ \forall A \subset\partial\Omega
  \\
  \int_A\Big[
  (\rho\,\bm{u}\cdot\bm{u} + p)\bm{n} - \dot P_0 \bm{n}
  \Big]\,dA=0 \ \ \ \ \ \forall A \subset\partial\Omega
  \\
  \int_A\Big[
  (E+p)\,\bm{u}\cdot\bm{n}   - \dot Q_0
  \Big]\,dA=0 \ \ \ \ \ \forall A \subset\partial\Omega.
\end{gather}\end{subequations}
Applying the fundamental lemma again shows that each of the integrands equate to zero everywhere on $\partial\Omega$.
Therefore, all three boundary conditions are recovered in the sharp interface limit.

\subsection{Essential and natural boundary conditions}

Since mass, momentum, and energy fluxes cannot be set simultaneously \cite{poinsot1992boundary}, it is generally not possible to prescribe boundary conditions in the form of \cref{eq:bndry_mass,eq:bndry_momentum,eq:bndry_energy}.
\replaced[id=R5,comment={5.3}]{Even when possible, it is generally inconvenient to specify boundary conditions in terms of fluxes; it is typically easier for them to be specified in terms of primitive variables.
}
{Rather,}
Boundary conditions are typically specified to be one of two possible varieties: ``essential,'' where velocity and density are prescribed but pressure is unknown; and ``natural,'' where pressure is prescribed but velocity and density are unknown.
In accordance with the partitioning of boundary condition types, the energy $E$ is split into its potential and kinetic components that are, in turn, expressed in terms of essential or natural variables,
\begin{align}
  E = K + U = \frac{1}{2}\rho\bm{u}\cdot\bm{u} + U(p).
\end{align}
Beginning by prescribing essential boundary conditions, the flux terms, \cref{eq:bndry_mass,eq:bndry_momentum,eq:bndry_energy}, are updated to include prescribed properties (denoted by the zero subscript), 
\begin{subequations}\begin{align}
  \dot{m}_0 &\mapsto \rho_0V_0 &  \forall\bm{x}&\in\partial_1\Omega\label{eq:bc_essential_mass}\\
  \dot{P}_0 &\mapsto \rho_0V^2_0 + p &  \forall\bm{x}&\in\partial_1\Omega \label{eq:bc_essential_momentum}\\
  \dot{Q}_0 &\mapsto \frac{1}{2}\rho_0V_0^3 + [U(p) + p](\bm{u}\cdot\bm{n}) &  \forall\bm{x}&\in\partial_1\Omega,\label{eq:bc_essential_energy}
\end{align}\end{subequations}
where $\partial_1\Omega$ is the essential boundary, $\rho_0:\partial_1\Omega\to\mathbb{R}$ is the prescribed boundary density, and $V_0:\partial_1\Omega\to\mathbb{R}$ is the prescribed boundary normal velocity.
Note that the non-prescribed quantities (such a pressure) are left unchanged, \added[id=R5,comment={5.3}]{meaning that they remain unknown quantities to be governed by the conservation equations, rather than by prescription}.
These modified fluxes are then substituted into the general formulation \cref{eq:diffuse_mass,eq:diffuse_momentum,eq:diffuse_energy}.
The prescribed values are left on the right hand side, but the unknown variables are then moved to the left hand side.
Application of the product rule yields the final result for the diffuse implementation of an essential boundary:
\cref{eq:bc_essential_mass,eq:bc_essential_momentum,eq:bc_essential_energy} may be substituted into \cref{eq:diffuse_mass,eq:diffuse_momentum,eq:diffuse_energy} to obtain
\begin{subequations}\begin{gather}
  \eta_{1\epsilon}\frac{\partial}{\partial t}(\rho) + \nabla\cdot[\eta_{1\epsilon}\rho\bm{u}] = \rho_0V_0|\nabla\eta_{1\epsilon}| \ \ \ \ \forall\bm{x}\in\mathbb{R}^3, \label{eq:diffuse_essential_mass}\\
  \eta_{1\epsilon}\frac{\partial}{\partial t}(\rho\bm{u}) + \nabla [\eta_{1\epsilon}(\rho\,\bm{u}\cdot\bm{u})] + \eta_{1\epsilon}\nabla p = \rho_0V_0^2 \nabla\eta_{1\epsilon} \ \ \ \ \forall\bm{x}\in\mathbb{R}^3, \label{eq:diffuse_essential_momentum}\\
  \eta_{1\epsilon}\frac{\partial}{\partial t}(E) + \nabla\cdot[\eta_{1\epsilon}(K)\bm{u}] + \eta_{1\epsilon}\nabla\cdot[(U+p)\bm{u}]= \frac{1}{2}\rho_0V_0^3|\nabla\eta_{1\epsilon}|  \ \ \ \ \forall\bm{x}\in\mathbb{R}^3. \label{eq:diffuse_essential_energy}
\end{gather}\end{subequations}
Note that subscript of $\eta_{1\epsilon}$ is used to indicate that the order parameter corresponds to an essential boundary in order to avoid confusion.
In the event that traditional boundaries are imposed along with diffuse boundaries, the corresponding boundary conditions would naturally be included here.
It is worth noting at this stage that this formulation circumvents the need to impose artificial or ``numerical'' boundary conditions; i.e., boundary conditions that do not drive the flow but are needed to complete the numerical solution.
(The imposition of a Neumann pressure boundary condition, for instance, is often needed for numerical reasons but is actually superfluous to the solution.)
Because the boundary is diffused, the unknown variables are simply allowed to follow the governing equations with no prescribed additional behavior.

The same procedure can be followed for natural boundary conditions, which specify the pressure but leave the velocity and density as unknowns (e.g., an outlet of a pipe).
As before, begin by updating the mass, momentum, and energy flux terms as follows,
\begin{subequations}\begin{align}
  \dot{m}_0 &\mapsto\rho\bm{u}\cdot\bm{n} &  \forall\bm{x}&\in\partial_2\Omega \label{eq:bc_natural_mass}\\
  \dot{P}_0 &\mapsto \rho\bm{u}\cdot\bm{u} + p_0 &  \forall\bm{x}&\in\partial_2\Omega \label{eq:bc_natural_momentum}\\
  \dot{Q}_0 &\mapsto \Big[\frac{1}{2}\rho\bm{u}\cdot\bm{u} + U(p_0) + p_0\Big](\bm{u}\cdot\bm{n}) &  \forall\bm{x}&\in\partial_2\Omega,\label{eq:bc_natural_energy}
\end{align}\end{subequations}
where $\partial_2\Omega$ denotes the natural boundary, and $p_0:\partial_2\Omega\to\mathbb{R}$ is the prescribed boundary pressure.
Again, the non-prescribed properties of velocity and density are left unchanged.
Substituting into the master set of equations, and using the subscript of $\eta_{2\epsilon}$ to describe an order parameter corresponding to a natural boundary, the following is found:
\begin{subequations}\begin{gather}
  \eta_{2\epsilon}\frac{\partial}{\partial t}(\rho) + \eta_{2\epsilon}\nabla\cdot[\rho\bm{u}] = 0 \ \ \ \ \forall\bm{x}\in\mathbb{R}^3\label{eq:diffuse_natural_mass}\\
  \eta_{2\epsilon}\frac{\partial}{\partial t}(\rho\bm{u}) + \eta_{2\epsilon}\nabla [\rho\,\bm{u}\cdot\bm{u}] + \nabla [\eta_{2\epsilon} p] = p_0 \nabla\eta_{2\epsilon} \ \ \ \ \forall\bm{x}\in\mathbb{R}^3, \label{eq:diffuse_natural_momentum}\\
  \eta_{2\epsilon}\frac{\partial}{\partial t}(E) + \eta_{2\epsilon}\nabla\cdot[ K \bm{u}] + \nabla\cdot[\eta_{2\epsilon}(U+p)\bm{u}] = [U(p_0)+p_0]|\nabla\eta_{2\epsilon}| \ \ \ \ \forall\bm{x}\in\mathbb{R}^3.\label{eq:diffuse_natural_energy}
\end{gather}\end{subequations}
As before, the unknown quantities are treated automatically, leaving the prescribed variables \added[id=R5,comment={5.3}]{(pressure and potential energy)} only to drive the flow.
It is important to note the differences between
\cref{eq:diffuse_natural_mass,eq:diffuse_natural_momentum,eq:diffuse_natural_energy} and 
\cref{eq:diffuse_essential_mass,eq:diffuse_essential_momentum,eq:diffuse_essential_energy}.
In addition to differing source terms, the location of the order parameter changes with respect to the derivative operators as a consequence of the presence of unknown boundary terms.
It is important to treat these terms correctly when numerically implementing this method; otherwise the boundary conditions will not be properly recovered.

This process may be used to derive diffuse boundary source terms corresponding to other boundary condition types.
One can also combine \cref{eq:diffuse_essential_mass,eq:diffuse_essential_momentum,eq:diffuse_essential_energy,eq:diffuse_natural_mass,eq:diffuse_natural_momentum,eq:diffuse_natural_energy} to a set of master equations for which both essential and natural boundary conditions are enforced by their corresponding order parameters.
Such a presentation is not included here, as it introduces complexity without any additional insight.
Moreover, caution should be taken in the implementation of such a combined boundary condition, as ill-defined behavior may occur at transition points and corners between boundary types.
(Corners, in particular, are in danger of violating the smoothness condition.)
Given that most applications of interest are unlikely to require multiple simultaneous types of diffuse boundary implementations, presentation and examples of a combined formulation are left to future work.

\subsection{Simplification to one-dimensional flow with gamma law EOS}

\replaced[id=RA]{Further important properties of this method will be demonstrated for }{The remainder of this work will be devoted to} the special case of one-dimensional flow.
The Euler equations are inviscid, and so there can be no tangential effect of the boundary on the flow.
Consequently, since only kinematic boundary conditions may be considered, all of the boundary effects are reducible to the one dimension aligned with the normal direction to the boundary.
\replaced[id=RA]{Working in two or three dimensions, while demonstrating the ease with which this method handles topological changes, does not affect the behavior or modelling of the boundary in the inviscid case.
The extension of this method to viscous flow will be left to future work.}{Two and three-dimensional results are therefore not relevant to this work, as fully resolved 2D\slash 3D examples conflate 1D boundary effects with normal flow behavior.
Such results would only serve to convey an overly optimistic view of this method.
Therefore, presentation of higher dimensional results, along with the extension of this method to viscous flow, will be left to future work.}

The following gamma-law EOS will be used for all subsequent analysis, which is simply
\begin{align}
  U = \frac{p}{\gamma-1}.
\end{align}
This statement is valid for a nonreacting, thermally perfect gas of constant heat capacity.
To further simplify the analysis, a single essential diffuse boundary is considered centered at some location $R(t)$, such that the right side is fluid and the left is solid.
The resulting governing equations, expressed in vector form, are
\begin{equation}
  \eta_\epsilon\frac{\partial}{\partial t}
  \begin{bmatrix}\rho \\ \rho u \\ K+U\end{bmatrix}
  +
  \frac{\partial}{\partial x}\eta_\epsilon
  \begin{bmatrix}\rho u \\ \rho u^2 \\ Ku\end{bmatrix}
  +
  \eta_\epsilon\frac{\partial}{\partial x}
  \begin{bmatrix}0 \\ p \\ (U+p)u\end{bmatrix}
  =
  \begin{bmatrix}\rho_0V_0 \\ \rho_0V_0^2  \\ \frac{1}{2}\rho_0V_0^3\end{bmatrix}\frac{\partial\eta_\epsilon}{\partial x}
  \ \ \ \ \ \forall x \in \mathbb{R} .
\end{equation}
The above also admits an equivalent quasi-linear form,
\begin{equation}
  \frac{\partial}{\partial t}
  \begin{bmatrix}
    \eta_\epsilon\rho \\ \eta_\epsilon u \\ p
  \end{bmatrix}
  +
  \begin{bmatrix}
    u & \rho & 0 \\ 0 & u & \eta_\epsilon/\rho \\ 0 & \gamma p / \eta_\epsilon &  u 
  \end{bmatrix}
  \frac{\partial}{\partial x}
  \begin{bmatrix}
    \eta_\epsilon\rho \\ \eta_\epsilon u \\ p
  \end{bmatrix}
  =
  \begin{bmatrix}
    \rho \\ u \\ 0
  \end{bmatrix}
  \frac{\partial\eta_\epsilon}{\partial t} +
  \begin{bmatrix}
    \rho_0V \\ V^2 \\ 0
  \end{bmatrix}
  \frac{\partial\eta_\epsilon}{\partial x}
  \ \ \ \ \ \ \forall x \in \operatorname{supp}\eta_\epsilon.\label{eq:1d_quasilinear}
\end{equation}
The quasi-linear form allows the eigenvalues of the coefficient matrix to be calculated:
\begin{align}
  \lambda_1 &= u - \sqrt{\frac{\eta_\epsilon}{\rho}\frac{\gamma p}{\eta_\epsilon}} = u-a
  &
    \lambda_2 &= u
  &
  \lambda_3 &= u + \sqrt{\frac{\eta_\epsilon}{\rho}\frac{\gamma p}{\eta_\epsilon}} = u+a,
\end{align}
which are identical to those of the unmodified Euler equations.
From this, it is shown that the speed of acoustic wave propagation is unchanged by the presence of a diffuse interface, even in the diffuse interface region.
The recovery of wave speeds in this diffuse method indicates compatibility with the  Baer-Nunziato model, as characteristic analysis has been used to compute the bidirectional coupling \cite{embid1992mathematical, andrianov2004riemann}. 
The continuous pressure field $p$ is distributed in the usual way.
In contrast, the quasi-linear form also makes it evident that the originally discontinuous fields $\rho$ and $u$, in the diffuse form, are distributed according to the order parameter.
With one caveat, to be discussed below, this distribution from non-fluid to fluid phase over the diffuse region guarantees a smooth transition.

Care must be taken, when considering the quasi-linear form, to apply it in the support of $\eta_\epsilon$ only (i.e., $x>R(t)-\epsilon$), because of the division by $\eta_\epsilon$.
There is no theoretical reason to explicitly consider the governing equations outside of this region, as they no longer hold outside of the domain.
It is computationally desirable, however, to avoid any explicit interface tracking.
Therefore, to adopt the above form everywhere, a small regularization is generally added to $\eta_\epsilon$ to render the quasi-linear form well-defined within the solid region.
The effect of this regularization, and the resulting impact on the fluid phase solution from the solid phase properties, will be discussed in the sections on implementation and results.

\section{Self-similarity and scaling}\label{sec:scaling}

In the application of the diffuse interface source term method, two facets of the modified equations are of concern.
First, it is essential to guarantee that the modified equations exhibit the same self-similar behavior as the Euler equations, so that comparisons may be made to classical analytic solutions for verification.
Second, and of greater consequence, it must be demonstrated that the introduction of a new diffuse boundary length scale $\epsilon$ does not affect the scale invariance of the flow.
If the flow away from the interface depends on $\epsilon$, this parameter may not be treated as a freely varying limit parameter, which we seek to avoid.

Scale invariant\slash self-similar solutions of the Euler equations are broadly used, as seen in \cite{velikovich2018generalized}.
Exhaustive discussions of their self-similar behavior can be found in the well-known works of Sedov \cite{sedov2018similarity}, Olver \cite{olver2000applications}, and many others.
More recently, the methods of Sophus Lie have been used to demonstrate that all self-similar behavior can be represented compactly as a Lie derivative, which generates the scaling group associated with the self-similarity \cite{barenblatt2003scaling}.
(For this reason, ``Lie derivative'' and ``scaling group generator'' are used interchangeably in this work.)
The infinitesimal symmetry approach of Lie analysis is perfectly suited to the present work, as all necessary smoothness is guaranteed, eliminating the need to resort to piecewise solutions (cf., free boundary hydrodynamic test problems discussed in \cite{ramsey2012guderley,ovsiannikov2014group,ze11966physics}).

This section makes use of exterior differential systems (EDS) and isovector analysis, which will be briefly reviewed below.
(Though necessary to establish the validity of the method, the following analysis in this section is technical and can be omitted by the reader without loss of continuity.)
For a thorough introduction to the isovector method, EDS, the underlying framework of exterior calculus, and the application to fluid dynamics, the reader is referred to \cite{frankel2011geometry,nakahara2018geometry,ramsey2017symmetries}.
For the present application, it will suffice to build on the following definitions.
{\it Differential n-forms} are fields (in some cases, similar to vector fields) that are integrable over an n-dimensional manifold.
{\it The wedge product} (or ``exterior product'') $\wedge$ is an anticommutative operator on differential forms, satisfying $\alpha\wedge\beta=-\beta\wedge\alpha$.
The result of a wedge product between an n-form and an m-form is an $\mathrm{n+m}$ form.
Finally, the {\it exterior derivative} $d$ is a general derivative on forms, and the action of the exterior derivative on an n-form $\alpha$ is an $\mathrm{(n+1)}$ form $d\alpha$, referred to as the differential of the form.
Like other derivatives, the exterior derivative is linear and obeys the chain and product rules.
It also possesses the important property that that $d^2=0$.

The application of isovector analysis applied to original Euler equations, with boundary conditions, is reviewed here, following the work of Ramsey \& Baty~\cite{ramsey2017symmetries} and Harrison \& Estabrook~\cite{harrison1971geometric}.
The one-dimensional Euler equations, with an idealized EOS, admit the EDS representation
\begin{subequations}\begin{gather}
  d\rho\wedge dx - u\,d\rho\wedge dt - \rho\,d u\wedge dt = 0 \ \ \ \ x>R \label{eq:euler_1d_mass}\\
  \rho\,du\wedge dx - u\,\rho\,du\wedge dt - dp\wedge dt = 0 \ \ \ \ x>R \label{eq:euler_1d_momentum}\\
  dp\wedge dx - u\,dp\wedge dt - \gamma\,p\,du\wedge dt = 0 \ \ \ \ x>R \label{eq:euler_1d_energy}\\
  \rho-\rho_0 = 0 \ \ \ \ x=R \label{eq:eulerbc_1d_density}\\
  u - V_0 = 0 \ \ \ \ x=R. \label{eq:eulerbc_1d_velocity}
\end{gather}\end{subequations}
The formulation of Euler equations as an EDS allows for straightforward derivation of the scaling group generator.
As shown in \cite{olver2000applications}, a point symmetry scaling group generator has the form
\begin{align}
  \mathcal{L}_v = \sum_i a_i\,x_i\,\frac{\partial}{\partial x_i},
\end{align}
where $a_i$ are constant scale factors (to be determined), and $\{x_1, x_2,\ldots\}$ is the vector of all independent and dependent variables.
It can be shown that $\mathcal{L}_v$ possesses the following properties: (1) $\mathcal{L}_v[x_i]=a_ix_i$; (2) $\mathcal{L}_v[dx_i] = d(\mathcal{L}_v[x_i])=a_i\,dx_i$; (3) $\mathcal{L}_v[f(x_i)] = a_ix_i\frac{\partial f}{\partial x}$ for $f$ an arbitrary Lipschitz function.

If a form of $\mathcal{L}_v$ can be found (that is, if constants $a_i$ can be determined) that returns zero when applied to the EDS, then the system is scale invariant.
Furthermore, the constants $a_i$ encode the scaling relationships between the independent and dependent variables.
Applying $\mathcal{L}_v$ to \cref{eq:euler_1d_mass,eq:euler_1d_momentum,eq:euler_1d_energy} and setting to zero yields the following group generator
\begin{align}
  \mathcal{L}_v = a_1x\frac{\partial}{\partial x} + a_2t\frac{\partial}{\partial t} + a_3\rho\frac{\partial}{\partial \rho} + (a_1-a_2)u\frac{\partial}{\partial u} + (a_3+2(a_1-a_2))p\frac{\partial}{\partial p}. \label{eq:generator_noboundary}
\end{align}
The three unitless parameters $a_1$, $a_2$, $a_3$ correspond to space\slash time scaling, velocity\slash pressure scaling, and pressure\slash density scaling, which all hold in the absence of a boundary.
If standard (non-diffuse) boundary conditions are present, they must also be invariant under $\mathcal{L}_v$.
Expressing the non-diffuse boundary conditions in EDS form, applying $\mathcal{L}_v$ to them and setting to zero induces the additional restrictions on $a_1$, $a_2$, $a_3$.
This calculation shows that the group generator reduces to
\begin{align}
  \mathcal{L}_v = a_1\,x\,\frac{\partial}{\partial x} + a_1\,t\,\frac{\partial}{\partial t}, \label{eq:generator_sharp_bndry}
\end{align}
in the presence of boundary conditions, \cref{eq:eulerbc_1d_density,eq:eulerbc_1d_velocity}.
The simplified form of \cref{eq:generator_sharp_bndry} compared to \cref{eq:generator_noboundary} reflects the lower symmetry of the system when subjected to boundary conditions.
The only remaining scaling behavior is the linear scaling of space with time, corresponding to the similarity variable $x/t$.

The diffuse boundary system must exhibit the same scaling behavior as the system with discrete boundary conditions.
Therefore, the task now is to demonstrate that the introduction of the characteristic length scale $\epsilon$, the diffuse interface width, does not alter the self-similar behavior of the flow.
Following the process described above, beginning with the EDS for the 1D, gamma-law diffuse boundary formulation:
\begin{subequations}\begin{align}
  -\eta\,d\rho\wedge dx + \eta\,u\,d\rho\wedge dt + \eta\,\rho\,du\wedge dt + (\rho u - \rho_0V_0)\frac{\partial \eta}{\partial x}dx\wedge\,dt &= 0 \label{eq:eds_diffuse_mass}\\
  -\eta\,\rho\,du\wedge dx + \eta\,\rho\,u\,du\wedge dt + \eta\,dp\wedge dt + (\rho u^2-\rho_0V^2)\frac{\partial \eta}{\partial x}dx\wedge dt &= 0 \label{eq:eds_diffuse_momentum}\\
  -\eta\,d\rho\wedge dx + \eta\,u\,dp\wedge dt + \gamma\,p\,\eta\,du\wedge dt + \frac{1}{2}(\rho u^3 - \rho_0V_0^3)\frac{\partial \eta}{\partial x}dx\wedge dt &= 0. \label{eq:eds_diffuse_energy}
\end{align}\end{subequations}
Because the boundary conditions are included implicitly as diffuse source terms, there is no need for an explicit boundary EDS.
Following the process described above, the group generator is determined by setting $\mathcal{L}_v=0$ for the above EDS and solving for the corresponding constants.
A straightforward but tedious calculation (see \ref{sec:scaling_algebra} for details) yields the important result
\begin{align}
  \mathcal{L}_v = b_1\,x\,\frac{\partial}{\partial x} + b_1\,t\,\frac{\partial}{\partial t} \ \ \ \ x \in \Omega_\epsilon,
\end{align}
for arbitrary constant $b_1$.
Apparently, this group generator is identical to that of the Euler system with boundary conditions \cref{eq:eulerbc_1d_density,eq:eulerbc_1d_velocity}.
This demonstrates that, despite the introduction of a diffuse lengthscale, linear scaling with space and time will still exist.
Therefore the diffuse interface system is consistent with the known effects of boundary conditions on the self-similar behavior of the flow.

Within the diffuse interface region, there is a functional dependence on $\eta$.
Recall that the only practical constraint on the form of $\eta$ is that it vary continuously from $0$ to $1$ across the thickness of the diffuse boundary.
It can be shown (\ref{sec:scaling_algebra}) that if $\eta$ is of the form $\eta=f(\xi)$ where $\xi=x/t$, that scaling will be preserved within the diffuse boundary region as well as $\Omega_\epsilon$
For example, if the interface thickness is set to be inversely proportional to time, i.e., $\epsilon\sim t^{-1}$, then it follows that any self-similar diffuse boundary solution must converge in time to the sharp interface limit.
It is generally unlikely that such a form for $\eta$ would be useful in practice, but this does suggest favorable scaling of the solution with decreasing $\epsilon$ for quasi-static flow.
For any $\eta$ that limits as required in \cref{sec:formulation}, the system is self-similar outside the support of $\eta$ for any interface width $\epsilon$.
This suggests that the refinement of the interface should be well-behaved for all realistic calculations; any spurious effects from the diffuse boundary will be limited to the diffuse region itself, and must vanish as the interface width goes to zero.

\section{Numerical implementation}\label{sec:implementation}

A Roe-averaged approximate Godunov solver implemented in python is used to solve the diffuse interface equations in one dimension.
A Godunov-type solver takes advantage of the equivalent wave speeds and eigenvectors between the Euler system and the diffuse interface system.
The primary difference between the two systems is the advected fields: in the diffuse interface system $\eta\rho$, $\eta u$, and $p$ are advected, compared to the standard Euler system where the fields $\rho$, $u$, and $p$ are advected.
The advected $\eta$ value is computed as the geometric average of the cell-centered $\eta$ values of neighboring cells, in keeping with the other Roe averages.
The characteristics of the diffuse interface equations are
\begin{align}
    d(\eta \rho) &- \eta\frac{dp}{a^2} &
    d(\eta u) &+ \eta \frac{dp}{a\rho} &
    d(\eta u) &- \eta \frac{dp}{a\rho};
\end{align}
as such, a factor of $\eta$ must be introduced to the Roe diffusion terms.
Furthermore, these factors of $\eta$ do not allow flux from the solid phase into the fluid phase, or vice-versa.
Rather, each phase interacts solely with the interface source terms; any transmitted properties are damped out or generated by the interface in a manner exactly equivalent to the transmission.
(This phenomenon becomes particularly apparent in the interaction of acoustic waves with the interface.)
For the numerical examples in this work, the order parameter is prescribed as
\begin{align}
    \eta(x,t) = \frac{1}{2}\left[\text{tanh}\left(\frac{x-R(t)}{4\epsilon}\right) + 1\right]
\end{align}
where $\epsilon$ is the interfacial width, and $R$ is the interface position, possibly varying in time with some prescribed velocity, i.e., $R=V(t)\,t$.
Again, the method should work for other choices of $\eta$.

Numerical issues arise in the implementation of the diffuse interface method outside of the fluid region and diffuse boundary region, due to the division by $\eta=0$.
In the context of a Godunov solver, the problem manifests in the calculation of sound speed outside the fluid region.
In order to compute a real sound speed everywhere in the domain, one practical solution is to couple to a stand-in solid phase such that the total density
\begin{align}
    \rho = (\eta - 1)\rho_{\text{solid}} + \eta\rho_{\text{fluid}}
\end{align}
is nonzero everywhere.
Thus the pseudo-solid phase has constant density, zero velocity, and spatially constant pressure everywhere.
Nevertheless, it does not contribute to the source terms in the interface, as all evolution equations are proportional to $\eta$.
Another way to mitigate this numerical difficulty is to implement a full coupling to the solid phase that provides physical meaning and evolution equations to the variables in the $\eta=0$ solid region; this is a rigorous approach that should be used for future work.

Although not explored in this work, the self-similar solutions discussed in \cref{sec:scaling} can provide benchmark solutions for additional code verification~\cite{moore2013self}.
It should also be noted that the artificial diffusion added to the Roe solver for stability does affect the self-similarity of the flow.
Fortunately, this effect is known to be small~\cite{velikovich2018generalized} and does not preclude the use of self-similar solutions for verification.

\section{Examples}\label{sec:examples}

\begin{figure}
  \centering
  \includegraphics[width=\linewidth]{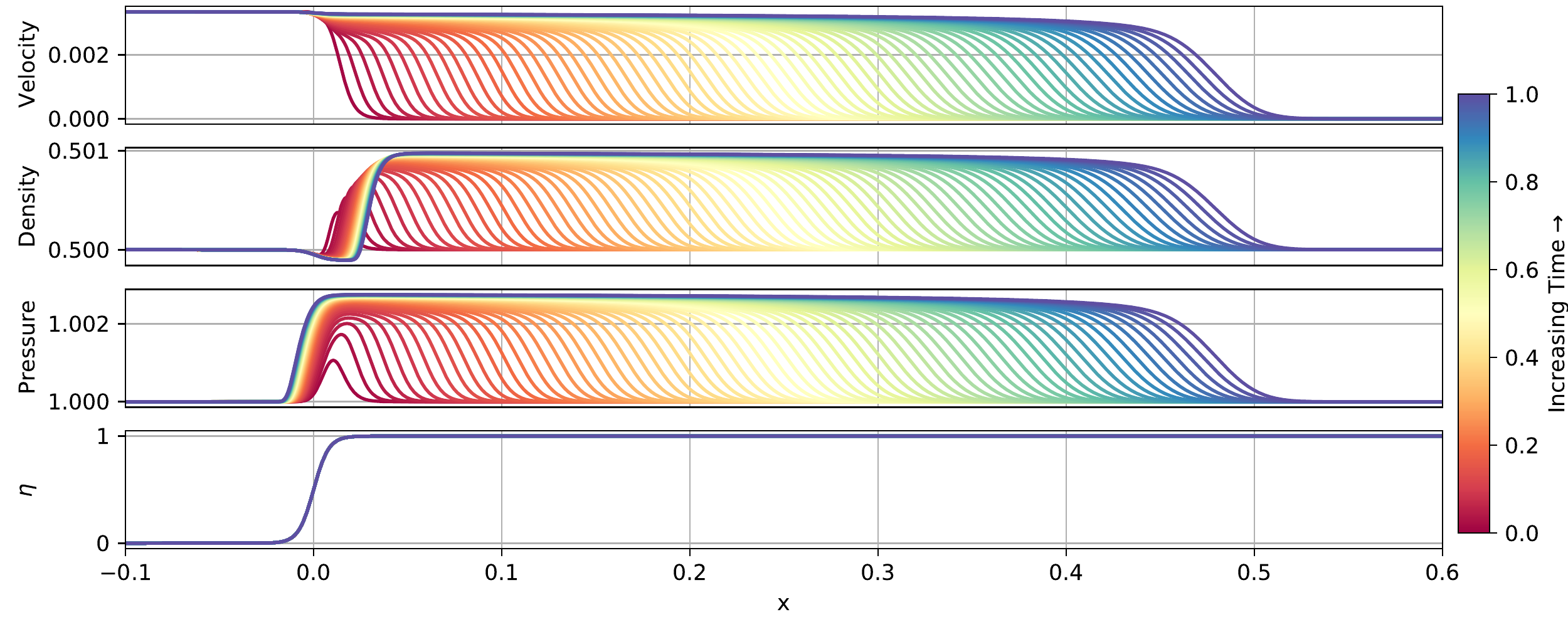}
  \caption{
    Numerical results: traces of velocity, density, pressure, and the (constant) order parameter $\eta$ at equal time intervals for a stationary boundary with prescribed mass flux.
    The flow is initialized with zero velocity and constant density and pressure.
    The positive mass flux at the boundary induces a pressure wave that propagates across the domain.
    The diffusivity of the flow is due to the artificial viscosity in the Riemann solver.
    Note the slight dip in density across the diffuse boundary.
    }
  \label{fig:NeumannV}
\end{figure}

\begin{figure}
  \centering
  \includegraphics[width=\linewidth]{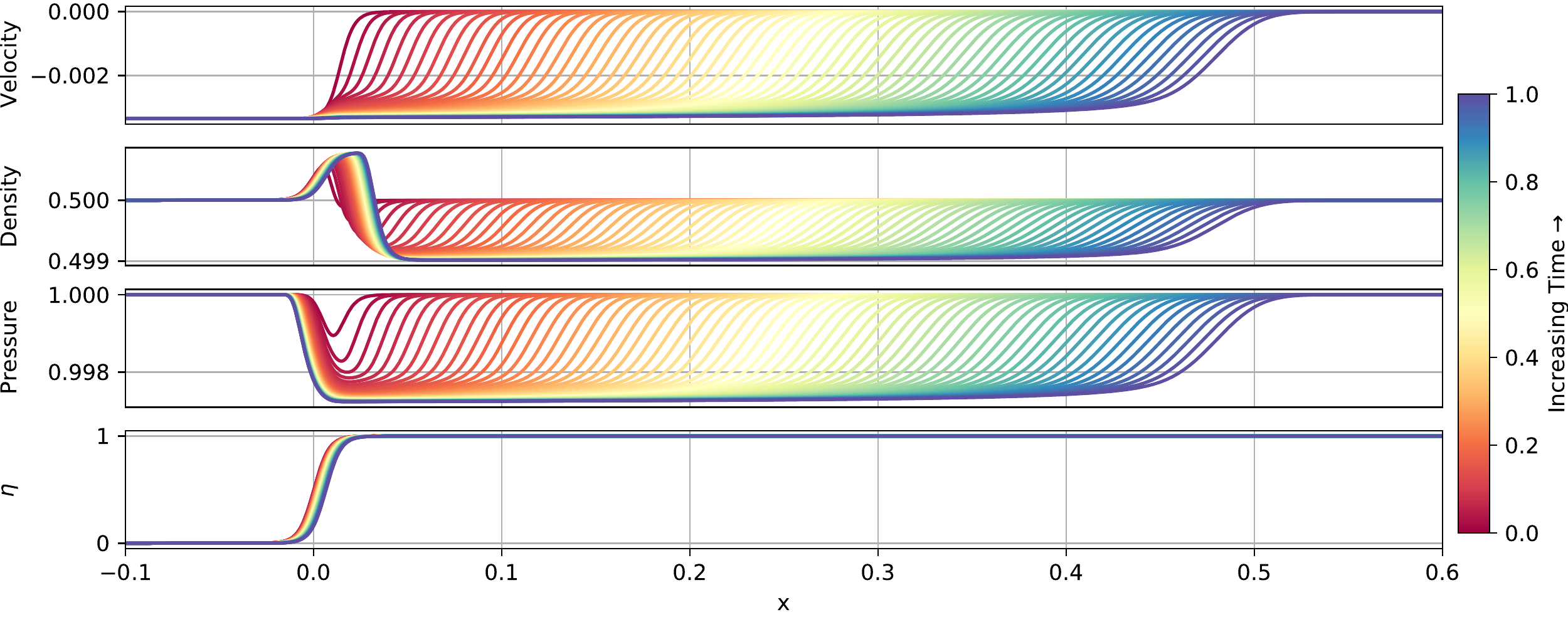}
  \caption{
    Numerical results: traces of velocity, density, pressure, and the order parameter $\eta$ at equal time intervals for a boundary moving from right to left.
    The flow is initialized with zero velocity and constant density and pressure.
    The boundary begins to move in the negative direction, resulting in an expansion wave.
    As in the previous case, the diffusivity of the flow is a result of artificial viscosity, and there is a slight density rise over the diffuse boundary region.
  }
  \label{fig:expansion}
\end{figure}

\begin{figure}
    \centering
    \includegraphics[width=\linewidth]{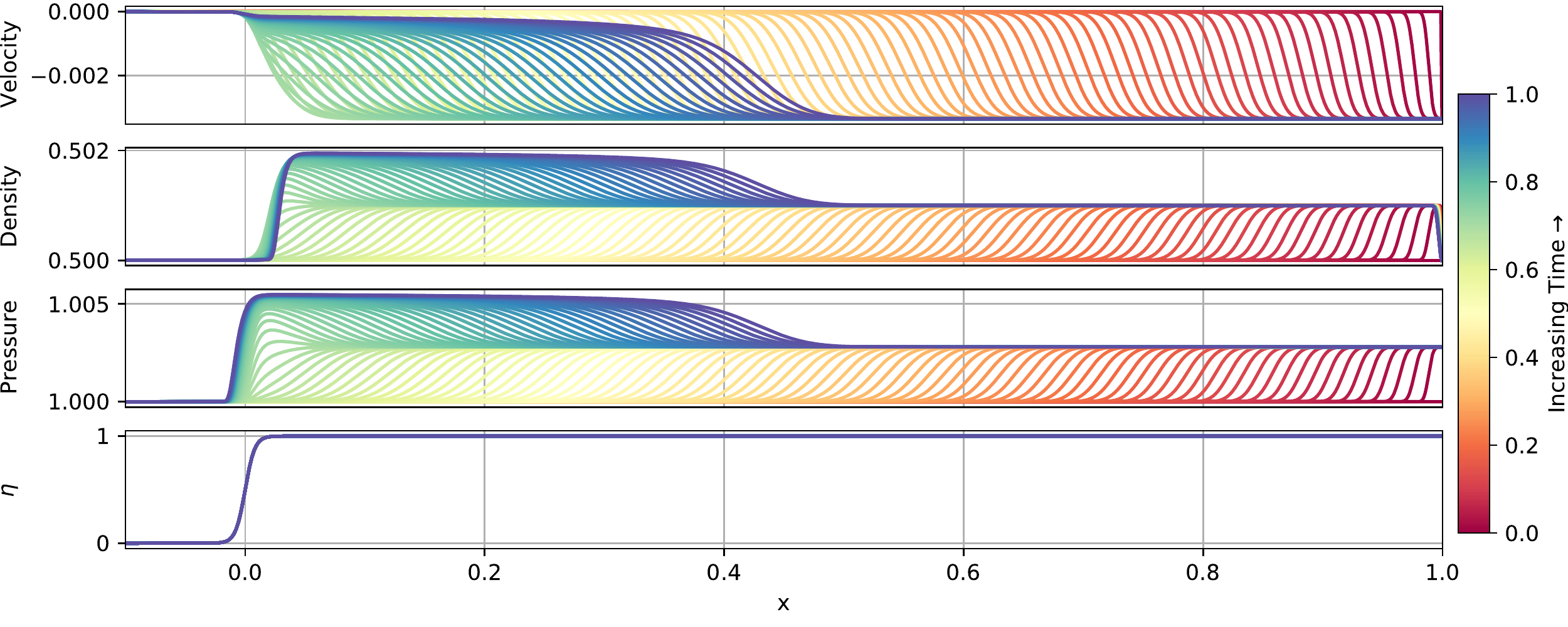}
    \caption{
        Numerical results: traces of velocity, density, pressure, and the order parameter $\eta$ at equal time intervals for a stationary boundary interacting with an incident wave.
        The flow is initialized to be zero in the domain, and a pressure wave is initiated at the right-hand side of the domain.
        The wave exhibits the expected behavior for a compression wave encountering a rigid boundary.
        Diffusivity is the result of artificial viscosity as before.
        Note the absence of the density rise in the diffuse region.
    }
    \label{fig:RefclectedWave}
\end{figure}

This section demonstrates the efficacy of the diffuse interface boundary method by means of a comprehensive set of numerical examples.
All results are generated using the methods described in the previous section.
The transient flow is represented using trace plots of velocity, density, and pressure with color corresponding to time.
It should be emphasized that the present results are for a boundary with a somewhat exaggerated width, having a diffusivity parameter of $\epsilon\approx0.05$, in order to clearly show the effects of the flow solver in the diffuse region and any far-field effects that result from it.
It should also be noted that all of the presented examples have been checked for numerical convergence, and are solved on a grid with sufficient resolution to guarantee arbitrarily high accuracy.
All convergence calculations reference an identical Roe Riemann solver on the gas-phase half of the domain.
A sharp boundary is placed at the interface location, and equivalent boundary conditions are set at that boundary. 

\subsection{Stationary boundary with mass flux}
\label{sec:injected}

In this example, a stationary boundary drives the flow by prescribing a mass (and corresponding momentum) flux across the diffuse interface.
Boundaries of this type are of central importance to problems involving deflagration.
The fluid domain is initialized at zero velocity and constant pressure and density.
At $t=0$, a mass flux boundary condition, in the form of an essential BC with prescribed velocity and density, is activated.
A wave is initiated at the boundary and propagates across the domain, eventually approaching fully developed flow (Figure~\ref{fig:NeumannV}).
The wave experiences some dispersion as a result of the artificial viscosity present in the Godunov solver; such dispersion is well-understood and is present regardless of whether a discrete or diffuse boundary is used.

The velocity profiles show that the flow does fully develop as expected.
It is apparent from the trace plots that the flow near to the boundary converges relatively slowly to the inlet velocity as compared the sharp boundary case.
In the density profile, a slight dip in the density is observed over the diffuse region, due to the incomplete coupling to the solid phase.
The absence of source terms due to the boundary conditions met by the solid manifests as irregular interface behavior on the left side of the diffuse interface.
The pressure field exhibits very little dependency on the diffuse boundary.
It is interesting to note that the derivative of pressure is approximately zero at the edge of the boundary, despite the fact that Neumann conditions for pressure are not explicitly imposed by the method.
This has the desirable effect of achieving a well-behaved, continuous pressure field through the interface and in the fluid region, which is useful when coupling to mechanics solvers in the solid region.

\subsection{Moving boundary}

In this example, the flow is driven not by an imposed mass flux but through the motion of the boundary itself, i.e., through a momentum flux.
This demonstrates the ability of diffuse interface methods to drive flow by immiscible interfaces, with potential application to solid-fluid interactions.
Here, the flow is initialized as with the previous example.
At time $t=0$, the interface begins moving from right to left (as can be seen in the trace plots of $\eta$), producing an expansion wave in the flow (Figure \ref{fig:expansion}).
The interface is moved slowly to ensure that the wave is acoustic, but no divergent behavior was observed for faster moving waves or after flow had fully developed.
As with the flow generated by a mass flux, this flow generated by a diffuse momentum flux boundary exhibits a slightly altered shape of the wave front influenced by the diffuse width and form of $\eta$.
As will be shown subsequently, however, this effect is small and vanishes in the sharp interface limit.

In general, the features of this flow are very similar to that of the mass-flux-driven case.
The main difference is in the density profile, which features an increase over the diffuse boundary rather than a decrease.
Again, however, this is a feature of the treatment of density in the solid phase, and can be mitigated through a complete solid-fluid coupling.

\subsection{Incident acoustic wave}

The final \added{one-dimensional} example demonstrates the passive interaction of the diffuse boundary with an incident acoustic wave.
Such a wave, upon encountering a perfectly rigid solid interface, will exhibit total reflection of the velocity (up to any numerical diffusion), and doubling of the pressure and density wave amplitudes.
This is observed in the diffuse boundary simulation (Figure \ref{fig:RefclectedWave}).
The diffuse boundary effect is pronounced in the reflected velocity wave, and can be seen by comparing the reflected wave traces to the incident wave traces.
Because there is no flux generated by the boundary, the density is observed to vary smoothly without the fluctuation visible in the prior two cases.
The pressure remains well-behaved across the interface region.

For this case, as with all of the test cases, convergence studies demonstrate that all deviations from the sharp interface solution vanish as $\epsilon\to0$.
Here, this results in a reflection coefficient of 1 for sufficiently small $\epsilon$.

\begin{figure}[h]
    \begin{subfigure}[c]{\textwidth}
        \includegraphics[height=1.5in,clip,trim=1.2cm 0.9cm 0.6cm .3cm]{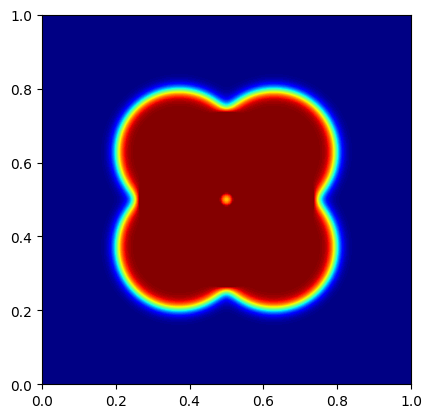}\hfill%
        \includegraphics[height=1.5in,clip,trim=1.2cm 0.9cm 0.6cm .3cm]{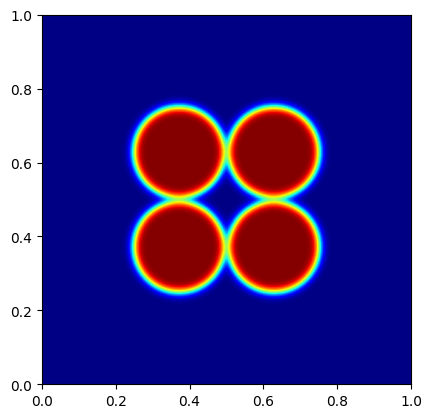}\hfill%
        \includegraphics[height=1.5in,clip,trim=1.2cm 0.9cm 0.6cm .3cm]{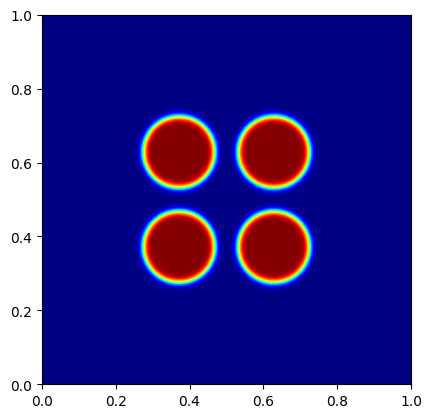}\hfill%
        \includegraphics[height=1.5in,clip,trim=1.2cm 0.9cm 0.6cm .3cm]{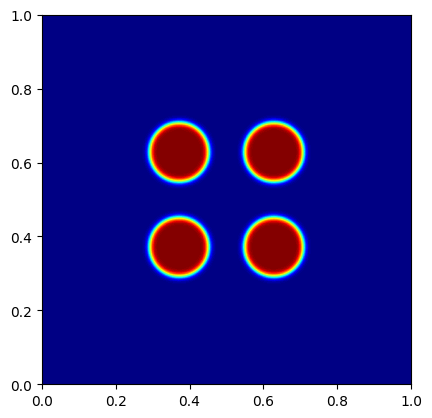}\hfill%
        \includegraphics[height=1.5in]{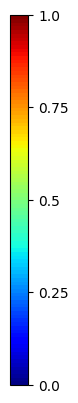}
        \caption{Order parameter $\eta\,[-]$}
        \label{fig:2D_results_eta}
    \end{subfigure}
    \begin{subfigure}[c]{\textwidth}
        \includegraphics[height=1.5in,clip,trim=1.2cm 0.9cm 0.6cm .3cm]{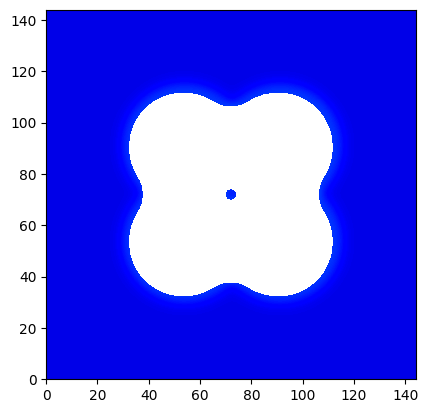}\hfill%
        \includegraphics[height=1.5in,clip,trim=1.2cm 0.9cm 0.6cm .3cm]{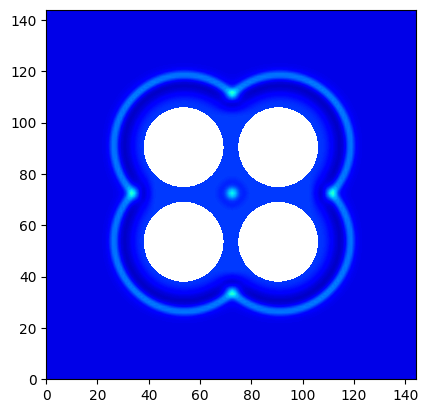}\hfill%
        \includegraphics[height=1.5in,clip,trim=1.2cm 0.9cm 0.6cm .3cm]{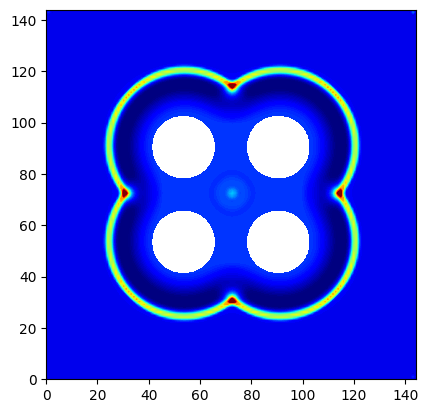}\hfill%
        \includegraphics[height=1.5in,clip,trim=1.2cm 0.9cm 0.6cm .3cm]{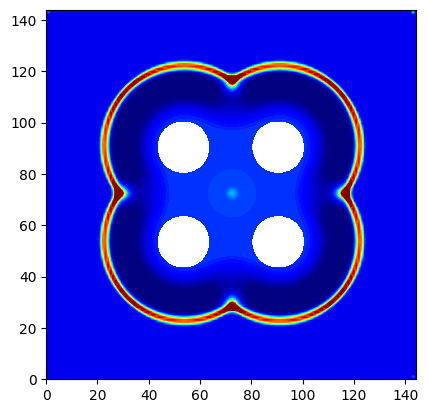}\hfill%
        \includegraphics[height=1.5in]{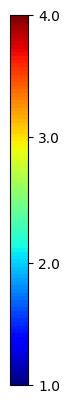}
        \caption{Density $\rho\,\mathrm{[kg/m^3]}$}
        \label{fig:2D_results_rho}
    \end{subfigure}
    \begin{subfigure}[c]{\textwidth}
        \includegraphics[height=1.5in,clip,trim=1.2cm 0.9cm 0.6cm .3cm]{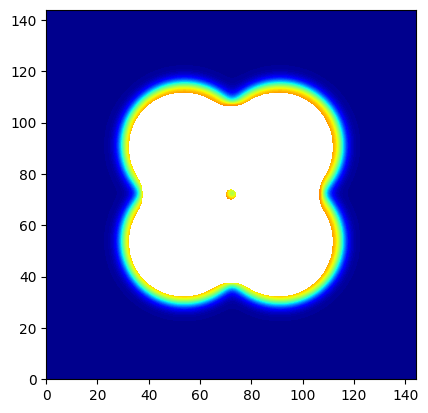}\hfill%
        \includegraphics[height=1.5in,clip,trim=1.2cm 0.9cm 0.6cm .3cm]{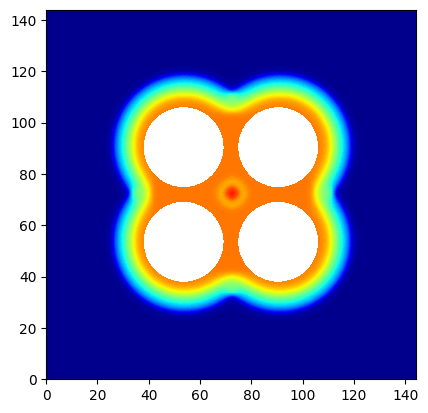}\hfill%
        \includegraphics[height=1.5in,clip,trim=1.2cm 0.9cm 0.6cm .3cm]{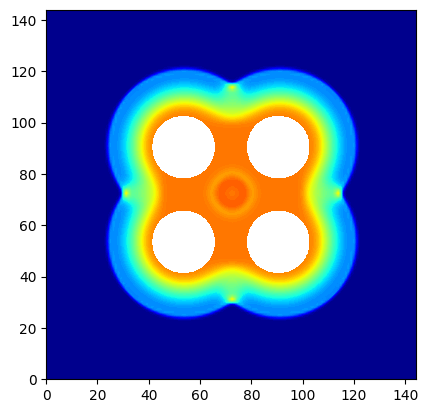}\hfill%
        \includegraphics[height=1.5in,clip,trim=1.2cm 0.9cm 0.6cm .3cm]{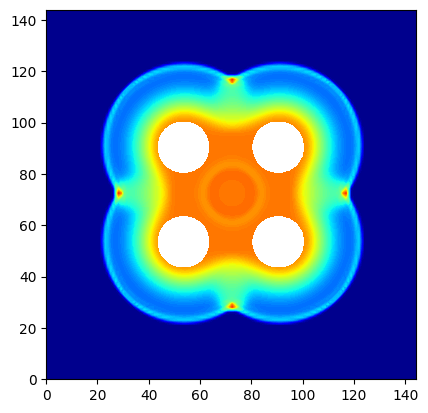}\hfill%
        \includegraphics[height=1.5in]{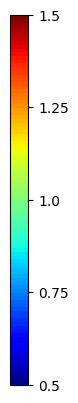}
        \caption{Pressure $p\,\mathrm{[Pa]}$}
        \label{fig:2D_results_p}
    \end{subfigure}
    \caption{\added[id=RA]{Flow induced by complex boundary evolution: a solid region defined by four shrinking, sublimating filled circles that initially overlap, then separate into four separate regions while effusing a mass flux. Topological transitions occur first at the emergence of a hypocycloid region in the center (first frame), followed by the separation into individual filled circles (second frame). 
    Images are shown at $t=0$, $0.01$, $0.02$, and $0.03\,\mathrm{s}$.
    The white regions reflect $\eta>0.9$.}}
    \label{fig:2d_results}
\end{figure}

\added[id=RA,comment={1.3,5.2}]{
\protect\subsection{Effusing, shrinking, overlapping filled circles}

This example demonstrates the effectiveness of the model in capturing complex, two-dimensional, boundary-driven flow in a case that would be difficult to capture using conventional simulation methods.
In this example, the flow is driven by a boundary that is simultaneously receding and effusing a prescribed mass flux.
The prescribed geometry can be seen in the plots of the order parameter $\eta$ (\cref{fig:2D_results_eta}).
The domain of the boundary is defined as the union of four overlapping filled circular regions, where each circle's center location is held constant, but the radius is steadily reduced in time.
As such, the domain undergoes two major topological changes.
First, the domain changes from simply-connected to toroidal, as an internal boundary around the hypocycloidal center region is created.
Second, the domain changes from toroidal to four distinct unconnected regions, as the overlapping filled circles separate.
For interface tracking methods, each of these transitions necessitates interface reconstruction, which is cumbersome and sometimes inaccurate.
For the diffuse boundary method, however, capturing this geometry and topological change requires no special treatment.
Here, the $\eta$ is simply prescribed, and the resulting flow is allowed to evolve naturally with no explicit interface handling.
The resolution of the grid is $324 \times 324$, ensuring that there are more than 16 grid points across the diffuse boundary to ensure convergence with respect to discretization.
The domain boundaries are given simple Dirichlet conditions for velocity, but the effects of the domain boundary are of no consequence to the present example over the time interval of interest.
Snapshots of density (\cref{fig:2D_results_rho}) and pressure (\cref{fig:2D_results_p}) indicate the evolution of the flow in time as calculated with the proposed method.

This example highlights the model's handling of other potentially problematic effects of topological transitions in a simulation.
In the present example, when surface separation occurs, there is a singularity in the flow due to the two newly-formed,  near-coincident fluxing boundaries.
% Here, the diffuseness of the interface acts as a regularization on flow singularities of this type, so that the resulting flow is smoothed according to the interface width parameter $\epsilon$.
% The effect of the singularity and its regularization can be seen in the second and third snapshots of density.
The effect of the singularity can be seen in the second and third snapshots of density, though the diffuseness of the method in the boundary prevents numerical issues.
In fact, here, the diffusiveness of the interface acts as a regularization on flow singularities of this type, so that the resulting flow is smoothed according to the interface width parameter $\epsilon$, which may be controlled by the user.
As a result, the numerical simulations handle the topological change seamlessly.
}

\subsection{Convergence}

\begin{figure}
  \includegraphics[width=\linewidth]{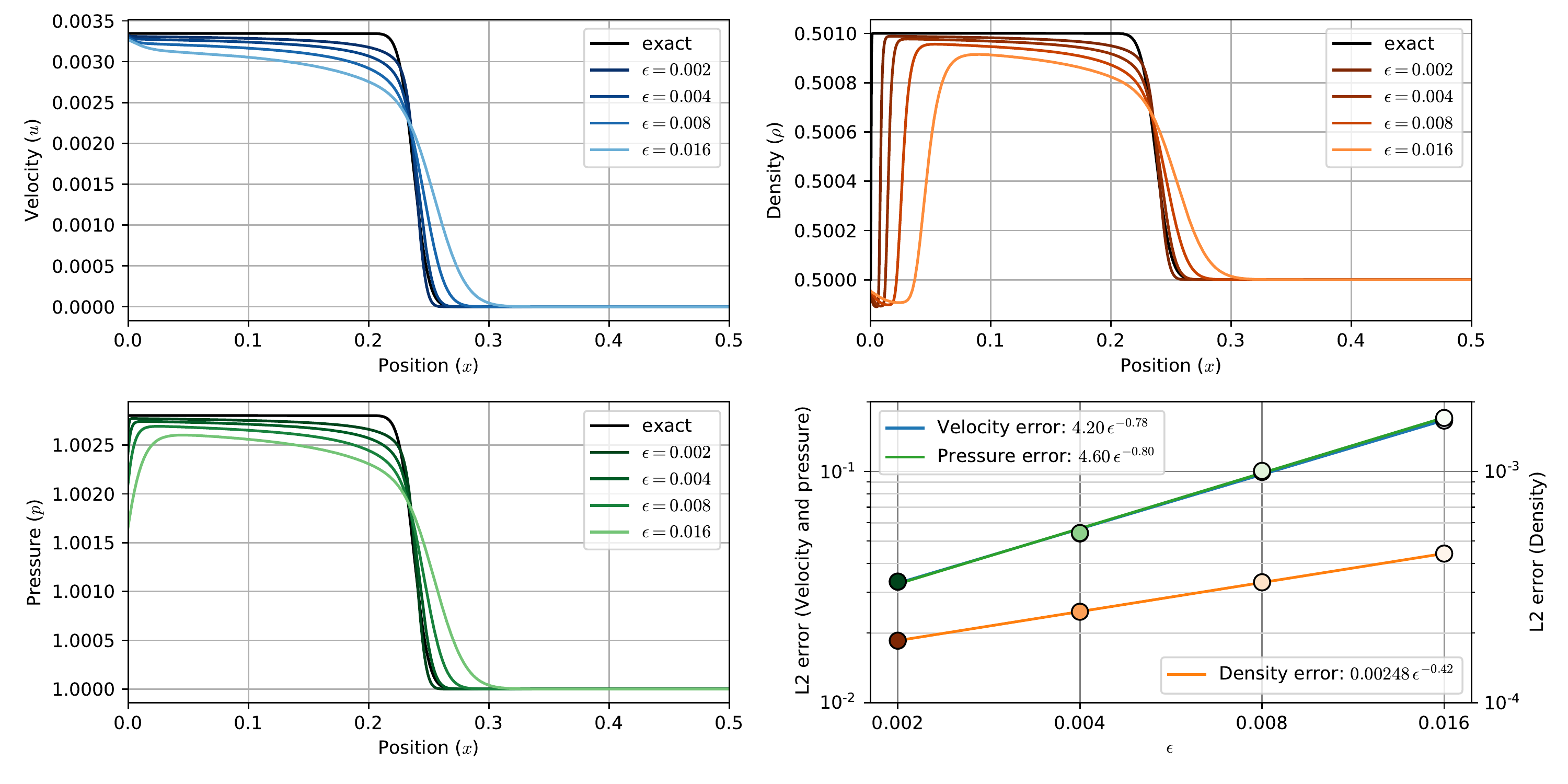}
  \caption{
    Convergence analysis of the diffuse interface method to the sharp (``exact'') result, with respect to diffuse boundary width and grid spacing, for a flow resulting from an imposed mass flux.
    Traces in the top two and lower left plots correspond to decreasing values of $\epsilon$, approaching the sharp interface solution.
    Errors are plotted in the lower right graph; note that pressure and velocity errors are calculated with respect to relative values (left axis) and density is calculated with respect to absolute values (right axis).
    Velocity error and pressure error are nearly coincident, and so only pressure error is visible.
  }
  \label{fig:convergence}
\end{figure}

This section demonstrates convergence of the method, with respect to the interface diffusiveness $\epsilon$, by comparing to the sharp interface solution.
All errors are measured with respect to the L2 norm.
The sharp interface solution is calculated using the same Roe Riemann solver with a sharp interface and equivalent boundary conditions, in order to isolate diffusive error due to the interface from diffusivity introduced by the Roe averages.

The most commonly cited disadvantage of diffuse interface methods is their computational cost, as diffuse interfaces must be fully resolved to avoid inducing excessive numerical error.
Though not explored here, judicious discretization methods, such as adaptive mesh refinement, are generally able to mitigate this computational expense.
All of the examples presented here are one dimensional, and it is therefore tractable to generate the simulation results by simply refining the domain everywhere to the maximum required resolution.
Multiple numerical convergence tests were run, showing that sufficient accuracy is reached when there are 16 grid points across the diffuse boundary.

\replaced[id=R1,comment={1.3}]{Quantitative convergence analysis was performed for all 1D examples, but}{Rather than showing and discussing each of the convergence studies}, for brevity, only one representative example is presented here.
\added[id=R1,comment={1.3}]{Additional convergence analysis is provided in \ref{sec:additional_verification}.}
The example considered is that of a flow generated by a mass flux imposed across the diffuse boundary, and allowed to evolve.
Boundary thicknesses of $\epsilon=0.016$, $0.008$, $0.004$, and $0.002$ are considered and compared against the sharp interface solution at $t=1$.
Qualitatively, it is clear that the traces for velocity, density, and pressure all approach the sharp interface solution as $\epsilon$ is decreased (\cref{fig:convergence} top left, top right, and lower left, respectively).
Interestingly, the average position of the wave appears to be independent of the value for $\epsilon$, as all of the traces cross at approximately $2/3$ of the propagating wave amplitude.
While it is not surprising that the waves would propagate at the same speed in the fully fluid domain, it is important that the diffusiveness of the boundary does not cause a delay in the compression wave propagation.

The L2 error is computed and plotted for velocity, density, and pressure (\cref{fig:convergence} lower right).
It is important to note that the errors in velocity and pressure were both normalized with respect to the relative velocity and the relative pressure, respectively, in the sharp interface solution.
On the other hand, the error in density is normalized with respect to the absolute density, which is a much larger value.
As a result, the absolute magnitude of the density error appears lower, even though it is not actually a closer match.
(In fact, the convergence rate of the density is actually lower than that of the velocity and pressure.)
Separate axes are used to more transparently compare the error behavior.

Quantitative analysis of the convergence indicates clearly that the error converges linearly.
Velocity and pressure both converge with an exponent of approximately $0.8$.
The density exponent is about half of that, approximately $0.4$.
This can be attributed to the density source term that (as discussed previously) induces a fluctuation within the diffuse interface.
It is clear, however, that the fluctuation is well-behaved and that its effect is significantly diminished for small values of $\epsilon$.
Therefore we concluded, based on this and many other tests, that the diffuse interface boundary method exhibits satisfactory convergence and exhibits good numerical behavior for all relevant diffuse boundary widths and boundary conditions.

\section{Conclusion}\label{sec:conclusion}

In this work we have presented a systematic method for reproducing arbitrary boundary conditions in Euler flow with diffuse source terms in a manner suitable for coupling to diffuse interface models such as phase field.
The time-varying diffuse boundary is regarded as an input to the fluid system, and no restrictions are made on the kind of boundary that can be imposed. 
We demonstrate that discrete boundary conditions can be replaced with diffuse boundary source terms that are guaranteed to be equivalent in the sharp interface limit.
The formulation is specialized to essential and natural boundary conditions here, although more exotic types of boundary conditions may be obtained by following the same process.
The remainder of the paper is restricted to one dimension, as all boundary effects in Euler flow are inherently normal to the boundary and not tangential.
The theoretical section of this paper concludes by addressing the essential question of scalability in the presence of a diffuse boundary.
We show, by recourse to exterior differential systems, that the scaling behavior of the Euler equations is entirely unaffected by the diffuse length scale.
Importantly, we further show that one can even expect self-similar scaling behavior within the diffuse boundary with respect to the similarity variable $x/t$.
Though it is unlikely that the solution within the diffuse boundary is ever of particular interest, this finding indicates that one can expect generally good behavior (i.e., one would not expect to see spurious oscillations) from a diffuse boundary.
The model is then demonstrated in the solution of a selection of pertinent boundary interaction problems, through the use of a Roe-averaged Godunov numerical solver.
We show that the diffuse boundary formulation accurately captures the flow resulting from an imposed mass flux, a regressing impenetrable boundary, and the interaction of an incident wave with the boundary.
More importantly we also demonstrate that the flow converges linearly in all variables of interest with respect to the sharp interface solution.

This work considers inviscid flow only, and extension to \deleted{two-dimensional and three-dimensional} viscous flows shall be left to future work.
%This work considers inviscid flow only, and extension to viscous flows shall be left to future work.
The generality of this formulation, however, lends itself immediately to a wide range of applications for which diffuse boundary motion models already exist, but for which the fluid is oversimplified or neglected altogether. 
The application of this work to systems of this nature may lead to new insights in the deflagration of solid composite propellants, growth of dendrites in batteries, or solidification of polycrystalline solids in casting processes.
Finally, though the focus of the current work has been on boundaries regularized for purely numerical reasons (i.e., with no physical meaning attached to the diffusivity), this model may further be applied to systems in which the transition region has physical significance.
Since the wave speeds are inherently preserved, the artificially thickened flame model is one particularly promising application.

\section*{Acknowledgements}
ES and BR acknowledge support from the Office of Naval Research, USA, grant number N00014-21-1-2113. 
This work used the INCLINE cluster at the University of Colorado Colorado Springs.
INCLINE is supported by the National Science Foundation, grant \#2017917.
Finally, the authors thank Dr.~Brian Bojko for his valuable insights on this work.

\section*{Declaration of Interests}
The authors have no conflicts to disclose.

\bibliographystyle{ieeetr} %jfm
\bibliography{main}

\appendix
\setcounter{figure}{0}

\section{Proof of \texorpdfstring{\cref{thm:boundary_integral}}{theorem 1}}\label{sec:thm_proof}

\begin{proof}
  The integrand in the theorem contains two parts, a zero-order term (multiplied by $\eta$) and a first-order term (multiplied by $\nabla\eta$).
  These will be treated separately.
  Beginning with the zeroth order term, it will now be shown that
  \begin{align}\label{eq:f_only}
    \lim_{\epsilon\to0}\int_A\int_{-\epsilon/2}^{\epsilon/2}f\eta\,ds\,d\bm{x} = 0 \ \ \ \forall A\subset\partial\Omega.
  \end{align}
  By applying the Cauchy-Schwartz inequality, and then the mean value theorem, it follows that
  \begin{align}
    \int_A\int_{-\epsilon/2}^{\epsilon/2}f\eta\,ds\,d\bm{x}
    \le \Bigg|\int_A\int_{-\epsilon/2}^{\epsilon/2}f\eta\,ds\,d\bm{x}\Bigg|
    \le \int_A\Bigg|\int_{-\epsilon/2}^{\epsilon/2}f\eta\,ds\Bigg|\,d\bm{x}
    \le \alpha|A|\epsilon,
  \end{align}
  for some non-negative finite constant $\alpha\in\mathbb{R}$.
  This follows from the boundedness of $f$ and $\eta$.
  Since the bounds of $f$ and $\eta$ do not depend on $\epsilon$ (unlike for $\nabla\eta$), $\alpha$ does not depend on $\epsilon$.
  It is then trivially true that
  \begin{align}
    - \Bigg|\int_A\int_{-\epsilon/2}^{\epsilon/2}f\eta\,ds\,d\bm{x}\Bigg| \ge -\alpha|A|\epsilon, 
  \end{align}
  which allows the integral to be bounded above and below:
  \begin{align}
    -\alpha |A|\epsilon \le \int_A\int_{-\epsilon/2}^{\epsilon/2}\Big[f\eta + g|\nabla\eta|\Big]ds\,d\bm{x}  \le \alpha|A|\epsilon.
  \end{align}
  Recalling that neither $\alpha$ nor $A$ depend on $\epsilon$, this shows that the integral reduces to zero in the limit at $\epsilon\to0$, proving (\ref{eq:f_only}).
  Now consider the second order part of the integral,
  \begin{align}
    \lim_{\epsilon\to0}\int_A\int_{-\epsilon/2}^{\epsilon/2}g|\nabla\eta|\,ds\,d\bm{x}.
  \end{align}
  Recall that the gradient of $\eta$ is always normal to the interface:
  \begin{align}
    |\nabla\eta| = \nabla\eta\cdot\bm{n} = \frac{\partial}{\partial s}\eta(\bm{x}+s\bm{n}).
  \end{align}
  Substituting in this form for $\nabla\eta$ and then integrating the inner (one-dimensional) integral by parts shows that
  \begin{align}
    \int_{-\epsilon/2}^{\epsilon/2}\Big(g(\bm{x}+s\bm{n})\frac{\partial}{\partial s}\eta(\bm{x}+s\bm{n})\Big)\,ds
    = g(\bm{x}+\bm{n}\epsilon/2) - \int_{-\epsilon/2}^{\epsilon/2}(\nabla g \cdot\bm{n})\eta\,ds.
  \end{align}
  Application of (\ref{eq:f_only}) shows that the second term vanishes in the limit as $\epsilon\to0$.
  The result is the first term, which when substituted back into the surface integral, yields
  \begin{align}
    \int_A\,g(\bm{x})\,d\bm{x}.
  \end{align}
  This concludes the proof.
\end{proof}

\section{Scaling Group Analysis of the Diffuse Interface System}
\label{sec:scaling_algebra}

This section provides additional steps in the derivation of the scaling group generator for the diffuse interface formulation. 
Begin by parameterizing a general scaling scaling group generator with constants $b_1\ldots b_5$ as
\begin{align}
 \mathcal{L}_v = b_1x\frac{\partial}{\partial x} + b_2t\frac{\partial}{\partial t} + b_3\rho\frac{\partial}{\partial \rho} + b_4u\frac{\partial}{\partial u} + 
 b_5p\frac{\partial}{\partial p}.
\end{align}
Application of $\mathcal{L}$ to the mass conservation component of the diffuse interface EDS (\cref{eq:eds_diffuse_mass}) and setting yields the following expression:
\begin{align}
  -(b_3+b_1)\eta d\rho\wedge dx + (b_4+b_3+b_2)(\eta u\,d\rho\wedge dt + \eta \rho\,du\wedge dt) + ((b_3+b_4+b_2+b_1)\rho u &\nonumber\\ - (b_1+b_2)\rho_0V_0)\frac{\partial \eta}{\partial x}dx\wedge\,dt 
  + (-d\rho\wedge dx + u\,d\rho\wedge dt &\nonumber\\+ \rho\,du\wedge dt) + (\rho u - \rho_0V_0)\left[b_1x\frac{\partial}{\partial x}\left(\frac{\partial \eta}{\partial x}\right) + b_2t\frac{\partial}{\partial t}\left(\frac{\partial \eta}{\partial x}\right)\right]dx\wedge\,dt &= 0.
\end{align}
Similar expressions can be found by applying the same process to \cref{eq:eds_diffuse_momentum,eq:eds_diffuse_energy}; this is left to the reader.
Restricting to the solution manifold through substitution~\cite{harrison1971geometric} results in
\begin{align}
  -(b_3+b_1)\left[\eta u\,d\rho\wedge dt + \eta \rho\,du\wedge dt + (\rho u - \rho_0V_0)\frac{\partial \eta}{\partial x}dx\wedge\,dt\right] &\nonumber\\ + (b_4+b_3+b_2)(\eta u\,d\rho\wedge dt + \eta \rho\,du\wedge dt) + ((b_3+b_4+b_2+b_1)\rho u - (b_1+b_2)\rho_0V_0)\frac{\partial \eta}{\partial x}dx\wedge\,dt &\nonumber\\
  + (-d\rho\wedge dx + u\,d\rho\wedge dt + \rho\,du\wedge dt) \left[b_1x\frac{\partial\eta}{\partial x} + b_2t\frac{\partial \eta}{\partial t}\right] & \nonumber\\ + (\rho u - \rho_0V_0)\left[b_1x\frac{\partial}{\partial x}\left(\frac{\partial \eta}{\partial x}\right) + b_2t\frac{\partial}{\partial t}\left(\frac{\partial \eta}{\partial x}\right)\right]dx\wedge\,dt &= 0,
\end{align}
along with similar expressions for momentum and energy.
Simplifying these and the above expressions yields the following restrictions on the scale factors $b_i$:
\begin{align}
 b_1 &= b_2  &
 b_3 &= 0 &
 b_4 &= 0 &
 b_5 &= 0.
\end{align}
Consequently, the scaling group generator reduces to 
\begin{align}
    \mathcal{L}_v = b_1\,x\,\frac{\partial}{\partial x} + b_2\,t\,\frac{\partial}{\partial t},
\end{align}
which is identical to that for the original Euler equations along with boundary conditions.
Note that the above holds in $\Omega_\epsilon$ only.
Within the interface, the additional constraint is placed on $\eta$ in order for self-similarity to hold:
\begin{align}
\mathcal{L}_v[\eta] = x\frac{\partial\eta}{\partial x} + t\frac{\partial \eta}{\partial t} &=0.
\end{align}
(In fact, this must hold everywhere, but it is trivially satisfied in the region $\Omega_\epsilon$ in which $\eta=1$.)
This differential equation admits a solution of the form
\begin{align}
    \eta(x,t) = f(x/t),
\end{align}
indicating that scaling holds within the interface if $\eta$ is a function of $x/t$.

\protect\section{\added[id=R1]{Supplemental convergence analysis}}\label{sec:additional_verification}

\begin{figure}[ht]
    \begin{subfigure}{\linewidth}
  \includegraphics[width=\linewidth]{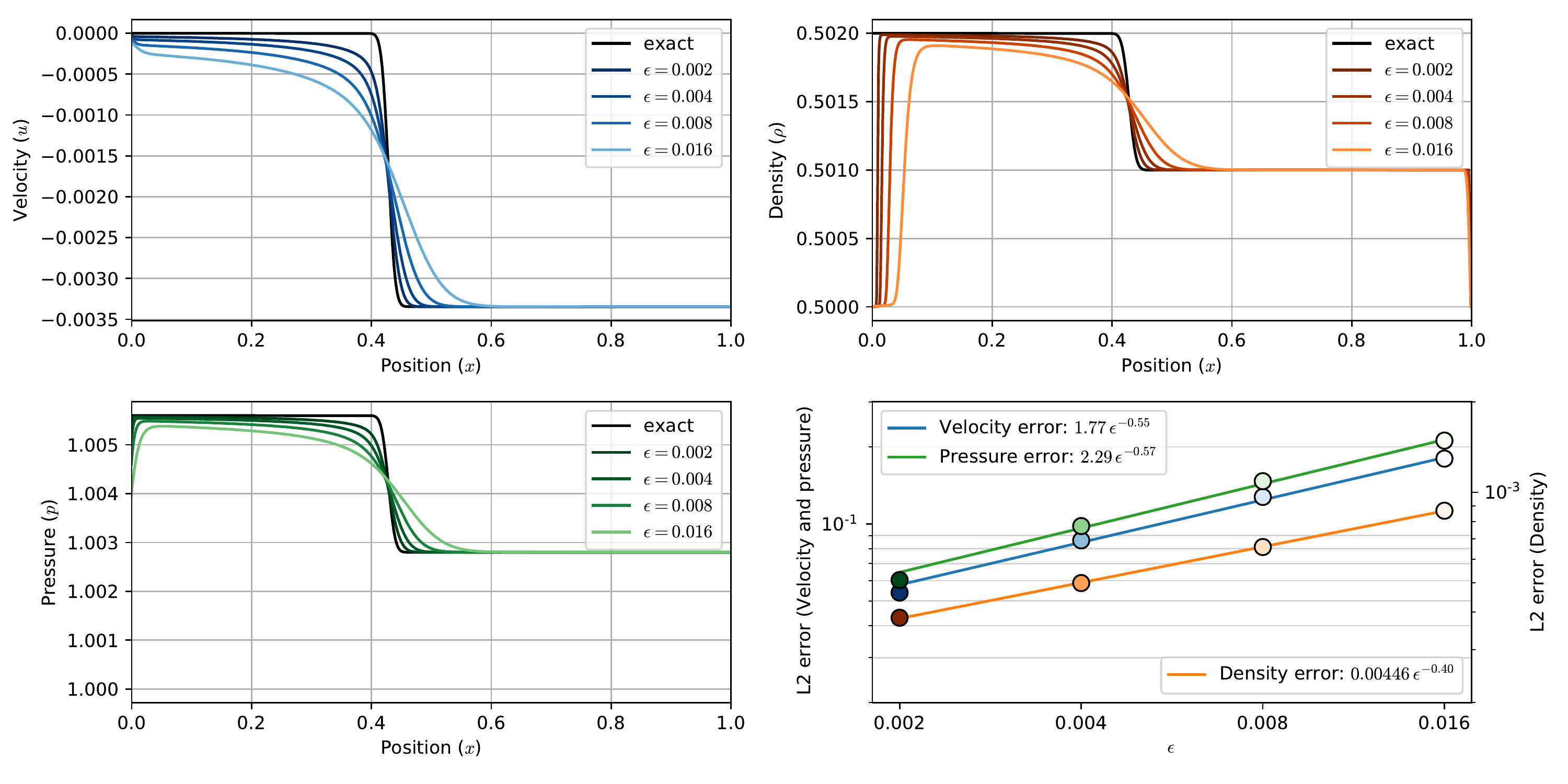}
  \caption{
    \added[id=R1]{Passive interaction of boundary with incident wave}
  }
  \label{fig:convergence_reflected}
  \end{subfigure}
    \begin{subfigure}{\linewidth}
  \includegraphics[width=\linewidth]{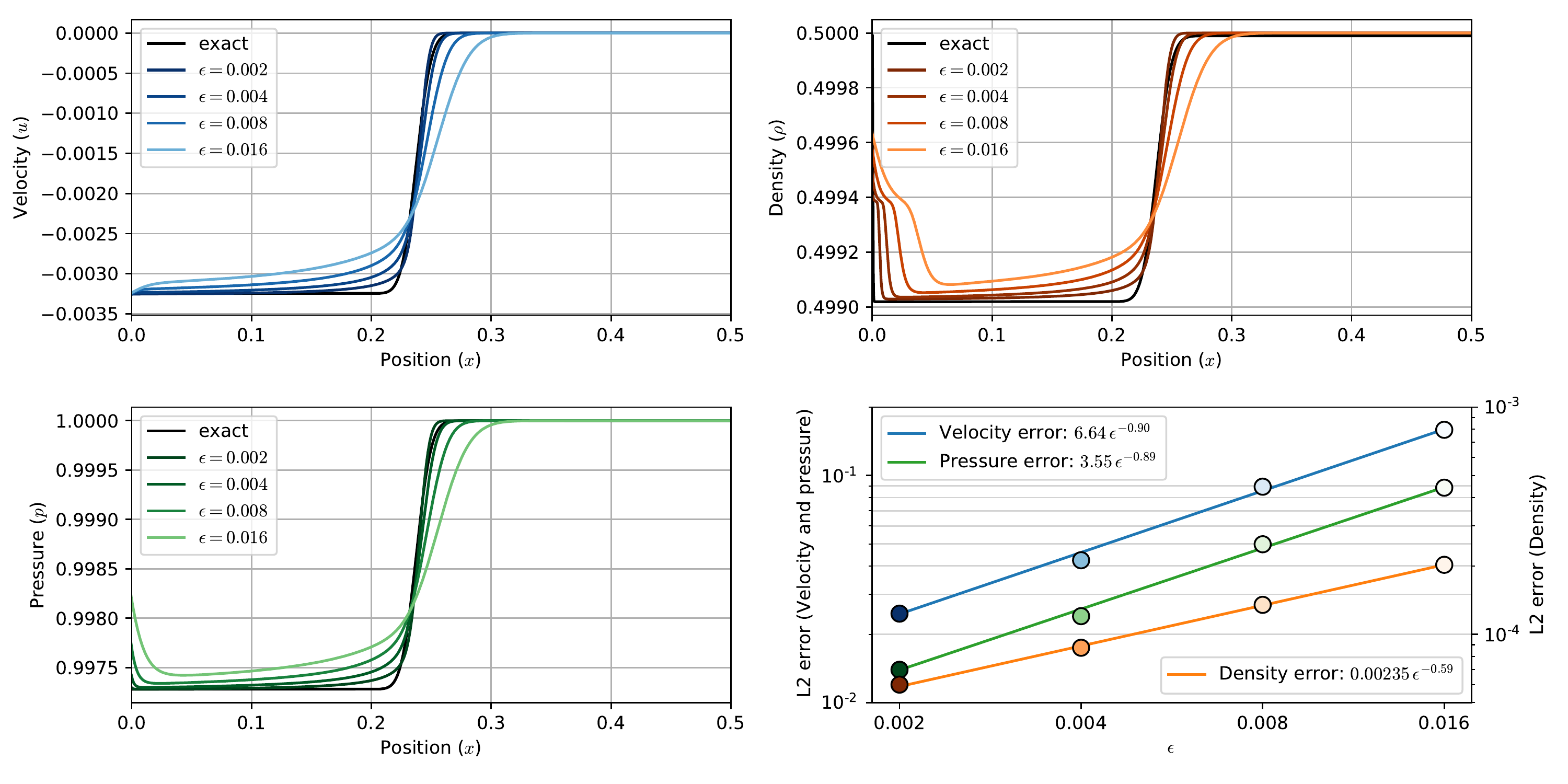}
  \caption{
    \added[id=R1]{Rarefaction wave induced by moving interface}
  }
  \label{fig:convergence_rarefaction}
  \end{subfigure}
  \caption{
      \added[id=R1]{Convergence analysis of the diffuse interface method to the sharp (``exact'') result, with respect to diffuse boundary width..
    Traces in the top two and lower left plots correspond to decreasing values of $\epsilon$, approaching the sharp interface solution.
    Errors are plotted in the lower right graph; note that pressure and velocity errors are calculated with respect to relative values (left axis) and density is calculated with respect to absolute values (right axis).}
    }
\end{figure}

\added[id=R1,comment={1.3}]{
This appendix presents the convergence data corresponding to the test cases of a passive wave interaction with a stationary boundary (\cref{fig:convergence_reflected}) and a rarefaction wave emitted from a moving interface (\cref{fig:convergence_rarefaction}).
The convergence benchmark is the sharp interface result.
In all cases, convergence with respect to spatial discretization was checked, so that the results reflect error induced by the diffuse boundary only.
As with the mass-flux driven example, the velocity, density, and pressure clearly exhibit linear convergence to the sharp-interface solution.
}

\end{document}